\documentclass[final]{elsarticle}

\usepackage[margin=1.25in]{geometry}

\usepackage{amssymb}
\usepackage{amsthm,amsmath,amssymb,braket,amsfonts}
\usepackage{xspace}
\usepackage{float}
\usepackage{etoolbox}
\usepackage{tikz}
\usepackage{algorithm}
\usepackage{algpseudocode}
\usepackage{mathtools}
\usepackage{cuted}
\usepackage{hyperref}
\usepackage[capitalise,nameinlink]{cleveref}
\crefname{equation}{}{}
\crefname{appendix}{}{}

\renewcommand{\vec}[1]{{\mathchoice
                {\mbox{\boldmath$\displaystyle{#1}$}}
                {\mbox{\boldmath$\textstyle{#1}$}}
                {\mbox{\boldmath$\scriptstyle{#1}$}}
                {\mbox{\boldmath$\scriptscriptstyle{#1}$}}}}
\newcommand{\mat}[1]{\mathbf{{#1}}}

\newcommand{\R}{\mathbb{R}}
\newcommand{\E}{\mathbb{E}}

\newcommand{\argmin}{\mathrm{argmin}}

\newcommand{\defeq}{\vcentcolon=}

\newcommand{\wxi}{w^{\vec\xi}}
\newcommand{\Exi}{\mathbb{E}^{\vec\xi}}

\newcommand{\pipostxi}{\pi_\text{post}^{\vec\xi}}
\newcommand{\piprxi}{\pi_\text{pr}^{\vec\xi}}
\newcommand{\pipostis}{\pi_\text{post}^\text{IS}}
\newcommand{\pipris}{\pi_\text{pr}^\text{IS}}
\newcommand{\Pis}{P^\text{IS}}
\newcommand{\Pxi}{P^{\vec\xi}}

\usepackage{lipsum}
\usepackage{subfigure}
\usepackage{graphicx}
\usepackage{epstopdf}

\ifpdf
  \DeclareGraphicsExtensions{.eps,.pdf,.png,.jpg}
\else
  \DeclareGraphicsExtensions{.eps}
\fi

\usepackage{enumitem}
\setlist[enumerate]{leftmargin=.5in}
\setlist[itemize]{leftmargin=.5in}

\journal{arXiv}

\begin{document}

\begin{frontmatter}



\title{Variance-based sensitivity of Bayesian inverse problems to the prior distribution
\tnoteref{thanks,funding}}
\tnotetext[thanks]{Submitted to the editors \today.}
\tnotetext[funding]{Supported in part by US National Science Foundation grants DMS \#1745654 and DMS \#1953271.}

\author[label1]{John Darges\corref{cor1}}\ead{jedarges@ncsu.edu}

\author[label1]{Alen Alexanderian}\ead{alexanderian@ncsu.edu}

\author[label1,label2]{Pierre Gremaud}\ead{gremaud@ncsu.edu}

\affiliation[label1]{organization={Department of Mathematics, North Carolina State University},
city={Raleigh},
state={NC},
 country={USA}}
 \affiliation[label2]{organization={The Graduate School, North Carolina State University},
city={Raleigh},
state={NC},
country={USA}}

\cortext[cor1]{Corresponding author}

\begin{abstract} 

The formulation of Bayesian inverse problems involves choosing prior
distributions; choices that seem equally reasonable may lead to significantly
different conclusions. We develop a computational approach to better
understand the impact of the hyperparameters defining the prior on the
posterior statistics of the quantities of interest. Our approach relies on
global sensitivity analysis (GSA) of Bayesian inverse problems with respect to
the hyperparameters defining the prior. This, however, is a challenging
problem---a naive double loop sampling approach would require running a prohibitive
number of Markov chain Monte Carlo (MCMC) sampling procedures. The present
work takes a foundational step in making such a sensitivity analysis practical
through (i) a judicious combination of efficient surrogate models and (ii) a
tailored importance sampling method. In particular, we can perform accurate
GSA of posterior prediction statistics with respect to prior hyperparameters
without having to repeat MCMC runs. We demonstrate the effectiveness of the
approach on a simple Bayesian linear inverse problem and a nonlinear inverse
problem governed by an epidemiological model.
\end{abstract}


\end{frontmatter}

\section{Introduction}\label{sec:intro}
Consider a Bayesian inverse problem governed by a system of differential
equations. The inverse problem uses a vector $\vec d$ of measurement data to
estimate the uncertain model parameters, $\vec \theta$.  The solution of the
Bayesian inverse problem is a posterior distribution $\pi_\mathrm{post}(\vec
\theta | \vec d)$.  After solving the inverse problem, typically we seek to make
some predictions based on the posterior. 
For example, for a prediction quantity $q(\vec \theta)$ we may consider 
\[
\mathbb{E}_\mathrm{post}(q) := \int q(\vec\theta)  \pi_\mathrm{post}(\vec\theta | \vec d)\,d\vec\theta. 
\]
A crucial component of this analysis is to know how the choice of prior
hyperparameters affects such predictions. We present a
practical variance-based global sensitivity analysis (GSA) approach to study how
statistics (e.g. mean or variance) of $q$ vary with respect to prior
hyperparameters. This enables us to identify which prior hyperparameters 
carry the most influence over the prediction.



Bayesian inference is pervasive; this perspective makes inferences not
just from data, but also by incorporating prior beliefs and assumptions. In
practice, these prior assumptions are often subjective choices made by
the researcher.  However, these prior beliefs can have a huge impact on the
results, including those of Bayesian inverse problems~\cite{Scales01}.   
This well-known issue motivated statisticians in the 1980s and 1990s
to develop a methodology, known as robust Bayesian
analysis~\cite{Berger90,Berger94,Hill94,Berger00}, for ensuring the robustness
of Bayesian inference to different choices by the researcher. These ideas have
continued to receive attention over the past two
decades~\cite{Lopes11,Brittleness,Watson16,Giordano18,BDA2,Nature21}.

\textbf{Related work.}
Sensitivity analysis of Bayesian inverse problems has been subject to several
recent research efforts.  
The articles~\cite{Sunseri22,Reese22,Chowdhary23} consider hyper-differential 
sensitivity analysis (HDSA) of Bayesian inverse problems.  HDSA is a technique
used originally for (deterministic) PDE-constrained optimization problems. 
HDSA, as a practical framework for sensitivity analysis of optimal control
problems governed by PDEs, was considered in~\cite{Hart20}.
In~\cite{Sunseri20}, HDSA was used for sensitivity analysis of deterministic
inverse problems to auxiliary model parameters and parameters specifying the
experimental setup (experimental parameters).  In~\cite{Sunseri22}, use of HDSA
is extended to nonlinear Bayesian inverse problems. Specifically, the authors 
consider the Bayes risk and the maximum a posterior probability (MAP) point as quantities of
interest for sensitivity analysis. 
In~\cite{Reese22}, the HDSA framework is used to study Bayesian inverse problems
governed by ice sheet models. The sensitivity of information gain, measured by
the Kullback--Leibler (KL) divergence between the prior and posterior, to
uncertain model parameters in linear Bayesian inverse problems is studied
in~\cite{Chowdhary23}.  HDSA provides valuable insight for experimenters on
where to focus resources during experimental design and when measuring auxiliary
parameters.  The previous works on HDSA of Bayesian inverse problems, have
focused primarily on sensitivity analysis with respect to auxiliary or
experimental rather than prior hyperparameters.  
More importantly, HDSA is local, relying on derivative information
 evaluated at a set of nominal parameters.  
Variance-based GSA, see~\cref{sec:gsa}, accounts for the uncertainty in the 
hyperparameters globally.  

The work~\cite{Vernon22}, which is closely related to our work,  examines
single-parameter statistical models using Bayesian inference.  In that paper,
the authors perform variance-based GSA on posterior statistics with respect to
prior and auxiliary hyperparameters.  Their method uses Gaussian process (GP)
surrogates to emulate the mapping from the hyperparameters to the posterior
distribution. This method requires many Markov chain Monte Carlo (MCMC)
runs to build the GP surrogate.  
For the Bayesian inverse problems we target, the high cost of evaluating the
forward model makes repeated MCMC runs impossible.  We
tackle this difficulty by using an importance sampling approach that allows
integrating the QoIs under study with respect to multiple posterior
distributions. Strategies for importance sampling on multiple
distributions have been subject to several previous works; see
e.g.,~\cite{GeyerThomp92,Madras99,GeyerThomp95,Geyer11,OwenIS}. We use
the structure of the Bayesian inverse problem to derive a tailored importance
sampling approach.  Another related work that has partly inspired the  approach
in the present work is~\cite{Merritt23}. That article, outlines a method
for GSA of rare event probabilities that combines surrogate-assisted GSA with
subset simulation.

\textbf{Our approach and contributions.}
We show that GSA is a viable computational approach to analyze the sensitivity of 
Bayesian inverse problems to prior hyperparameters.
The proposed approach is goal oriented---the focus is on the posterior 
statistics of prediction/goal QoIs that are functions of the inversion parameters.  
We first frame the problem in a
manner conducive to variance-based GSA in~\cref{sec:setup}. We detail the
computational strategy for sensitivity analysis in~\cref{sec:method}. Our method
combines two key techniques. Importance sampling eliminates the need for
repeated MCMC runs for different choices of the prior. Then, sparse polynomial
chaos expansion (PCE) and extreme learning machine (ELM) surrogate models emulate
the mapping from prior hyperparameters to statistics of $q$.  Use of surrogate
models not only eases the computational burden, but also improves the accuracy
of the sensitivity analysis. The combined approach enables prior hyperparameter sensitivity
analysis for many Bayesian inverse problems. If one has access to a single MCMC
run, then one can ascertain prior hyperparameter importance.  
To demonstrate the effectiveness of the proposed approach, we present extensive
computational experiments in the context of two examples: a simple linear
inverse problem in Section~\ref{sec:linear} and a nonlinear inverse problem
governed by an epidemiological model in Section~\ref{sec:seir}. 



\section{Hyperparameter-to-statistic mapping of Bayesian inverse problems}\label{sec:setup}
In an inverse problem~\cite{InvProb}, we use a model and observed data to
estimate unknown model parameters of interest.  We 
consider the inverse problem of estimating a parameter vector 
$\vec \theta$ in models of the form
\begin{equation}\label{equ:model}
\left\{\begin{aligned}
&\vec y' = f(\vec y;\vec\theta),\\
&\vec y(t_0)=\vec y_0.
\end{aligned}  \right. 
\end{equation}
Here, $\vec y \in \R^d$ is the state vector.  In a deterministic formulation of
the inverse problem, we typically seek a $\vec\theta$ that minimizes the cost
functional,
\begin{equation}\label{equ:cost}
J(\vec\theta):=\|\mat{B} \vec y(\vec\theta)-\vec{d}\|^2.
\end{equation}
Here, $\vec d$ is a vector of data measurements, 
$\mat{B}$ is a linear operator that selects the corresponding model
responses, and $\vec y$ is obtained by solving~\eqref{equ:model}. 



We focus on Bayesian inverse problems~\cite{InvProb} and seek a statistical distribution for $\vec \theta$, known as the posterior
distribution, that is conditioned on the observed data and is
consistent with the prior distribution. In this context, the prior
distribution encodes our prior knowledge 
regarding the parameters. The Bayes formula 
shows how the model, data, and the prior are combined to 
obtain the posterior distribution:
\begin{equation}\label{equ:bayes}
    \pi_\mathrm{post}(\vec\theta|\vec
d)\propto \pi_\mathrm{like}(\vec d|\vec{\theta})\times
\pi_\mathrm{pr}(\vec{\theta}),
\end{equation}
where $\pi_\mathrm{like}$ is the data likelihood and $\pi_\mathrm{pr}$ is the
prior probability density function (PDF). Throughout this paper, we assume a Gaussian noise 
model for the observation error. In this case, the Bayes formula reads
\begin{equation}\label{equ:gbayes}
 \pi_\mathrm{post}(\vec\theta|\vec
d)\propto \text{exp}\big({-\frac{1}{2}(\textbf{B} \vec y(\vec\theta)-\vec{d})^\top\Gamma_\mathrm{noise}^{-1}(\textbf{B} \vec y(\vec\theta)-\vec{d})}\big)\times\pi_\mathrm{pr}(\vec{\theta}) ,
\end{equation}
where $\Gamma_\mathrm{noise}^{-1}$ is the noise covariance.

In practice, we are often interested in scalar prediction quantities of interest
(QoIs) that depend on $\vec \theta$. Let $q(\vec \theta)$ be such a QoI.  
Solving the Bayesian inverse problem enables reducing the uncertainty in $\vec\theta$ and
consequently in $q(\vec\theta)$. In this case, the statistical properties of $q$
depend on $\pi_\mathrm{post}$.  Let $\Psi(q)$ denote a generic statistic of $q$.
Examples include $\Psi(q)=\mathrm{var}(q)$ or $\Psi(q)=\mathbb{E}(q)$,
where the expectation and variance are with respect to the posterior
distribution. Another example is $\Psi(q)=q(\vec\theta_\mathrm{MAP})$; i.e, QoI evaluated 
at the maximum a posteriori (MAP) point estimate of $\vec\theta$. 
Recall that the MAP point, $\vec\theta_\mathrm{MAP}$, is 
a point where the posterior PDF attains its maximum value. Using the Bayes formula~\eqref{equ:gbayes}, we note that the MAP point is the solution to the nonlinear least squares problem
\begin{equation}\label{equ:mappoint}
\vec\theta_\mathrm{MAP} = \underset{\vec\theta}{\argmin} \; 
J(\vec\theta) \defeq (\textbf{B} \vec y(\vec\theta)-\vec{d})^\top\Gamma_\mathrm{noise}^{-1}(\textbf{B} \vec y(\vec\theta)-\vec{d}) - 2 \log(\pi_\mathrm{pr}(\vec\theta)).
\end{equation}

We consider how the choice of prior affects
$\Psi(q)$. Narrowing this question, we take a parameterized family
of prior distributions $\pi_\mathrm{pr}^{\vec\xi}(\vec\theta)$ determined by a
vector $\vec\xi$ of scalar hyperparameters. For a
Gaussian prior, the hyperparameters can be taken as the prior means and
variances. With this setup, the choice of
$\vec\xi$ will determine our statistic of interest so that 
$\Psi(q) = \Psi^{\vec\xi}(q)$. 
In what follows,
the \textit{hyperparameter-to-statistic (HS) mapping} $F:\R^n\to\R$ is given by
\begin{equation}\label{equ:hsmap}
F(\vec\xi) \defeq \Psi^{\vec\xi}(q).
\end{equation}
To model the uncertainty in the hyperparameters, we consider them as random
variables and then analyze  how the uncertainty in the entries
of $\vec\xi$ contributes to the uncertainty in $F(\vec \xi)$.  To this end, we
follow a variance-based sensitivity analysis framework, and compute the Sobol' indices~\cite{SalSob95,Sobol01} of the HS mapping $F$ 
with respect to $\vec\xi$.

 
For the purposes of this study, we let the prior hyperparameters $\vec\xi$
follow uniform distributions, $\xi_j\sim\mathcal{U}[a_j,b_j]$, for
$j=1,\ldots,n$. We focus on three choices for the statistic of interest $\Psi$
in~\eqref{equ:hsmap}:
\begin{itemize} 
\item the mean: $F_\mathrm{mean}(\boldsymbol\xi) = \mathbb{E}^{\vec\xi}_\mathrm{post}(q)$;
\item the variance: 
$F_\mathrm{var}(\vec\xi) = \mathbb{E}^{\vec\xi}_\mathrm{post}(q^2) - 
(\mathbb{E}^{\vec\xi}_\mathrm{post}(q))^2$; and
\item the QoI evaluated at the MAP point: $F_\mathrm{MAP}(\vec\xi) \defeq q(\vec\theta_\text{MAP}(\vec\xi))$, with 
$\vec\theta_\text{MAP}(\vec\xi)$ from~\eqref{equ:mappoint}.
\end{itemize}
The mean and variance are computed from moments of the posterior PDF.  These two
quantities can be estimated at each $\vec\xi$ by Monte Carlo integration.  Estimating $F_\text{MAP}$ instead requires solving the nonlinear least
squares problem~\eqref{equ:mappoint} for each $\vec\xi$.

\section{Global sensitivity analysis and surrogate-assisted approaches}\label{sec:gsa}
We focus on variance-based
GSA using Sobol'
indices~\cite{SalSob95,Sobol01,Saltelli08,Smith14,IoossLeMaitre15,PrieurTarantola17}. 
Consider a (scalar-valued) model
\[
y = F(\vec x), \quad \vec x \in \R^d.    
\]
We assume that the components of $\vec x$
are independent random variables.  
In variance-based GSA, the most important inputs are those that
contribute the most to the output variance $\mathrm{var}(F(\boldsymbol x))$.
Sobol' indices are quantitative measures of this contribution.
Specifically, 
the first-order Sobol' indices, $S_k$, and the total Sobol' indices
$S^\mathrm{tot}_k$, are defined by 
\begin{equation}\label{equ:sobol}
S_k = \frac{\mathrm{var}(F_k)}{\mathrm{var}(F)},\quad S_k^\mathrm{tot} = 1 - \frac{\mathrm{var}(\mathbb{E}(F|x_l,l\neq k ))}{\mathrm{var}(F)},
\end{equation}
where $F_k(x_k)\defeq\E(f|x_k) - \E(f)$. In practice, the Sobol' indices are
approximated by Monte Carlo sampling, requiring many evaluations of the
model~\cite{Saltelli10}.  This can be too costly, especially when the model $F$
is expensive to evaluate.  In such cases, it is common practice to construct a
surrogate model $\widehat F\approx F$ whose Sobol' indices can be efficiently
computed~\cite{Sargsyan2017,GratietMarelliSudret17}. In the best case scenario,
the Sobol' indices of the surrogate model can be computed analytically.  We
detail two such surrogate models below. 

\textbf{Polynomial chaos surrogates.}
Polynomial chaos expansions (PCEs) take advantage of orthogonal polynomials to
approximate expensive-to-evaluate models; see~\cite{Sudret08,Crestaux09}.
The standard approach is to truncate the PCE 
based on the total polynomial degree. 
%
%
PCE surrogates are advantageous because they admit analytic formulas for Sobol'
indices that depend only on the PC coefficients~\cite{Sudret08}.  In practice,
the PC coefficients are typically computed using non-intrusive approaches that
involve sampling the model $F$. These  include non-intrusive spectral projection
or regression based methods~\cite{LeMaitreKnio10}.  In the present work, we build PCE surrogates using
sparse regression~\cite{Blatman10,Blatman11}. As noted in~\cite{Merritt23}, this
approach is particularly useful in the case where function evaluations are
noisy.  Solving the sparse regression problem can be formulated as a linear
least squares problem regularized by an
$\ell^1$-penalty~\cite{Doostan17,Marelli17}.  In our numerical computations, we
use the SPGL1 solver~\cite{BergFriedlander08,spgl1site} to solve such problems. 
Note that an $\ell^1-$penalty approach also involves choosing a penalty
parameter.  In our experiments, we use PCE surrogates with the basis truncated at total
degree $5$, and we perform a tenfold cross validation over training sets to
choose the $\ell^1$-penalty parameters.

\textbf{Sparse weight-ELM surrogates.} 
Sparse weight extreme learning machines (SW-ELMs) are a class of neural network surrogates that build on the standard extreme learning machines (ELMs). They are single-layer neural networks of the form
\begin{equation}\label{equ:elm}
\widehat F(\vec x) = \vec\beta^\top \phi(\mat W\vec x + \vec b),\quad\vec x\in\R^d.
\end{equation}
Here, $\vec\beta$ denotes the output weight vector, $\mat W$ the hidden layer weight matrix, $\vec b$ the hidden layer bias vector, and $\phi$ the activation function. The weights and biases are usually trained all at once by solving a nonlinear least squares problem. 
ELMs instead use randomly chosen hidden layer weights and biases.  Training an ELM then only
involves determining the output layer weights by  solving a linear-least squares
problem; see~\cite{HuangELM,Huang11} for details.  SW-ELM~\cite{ELM} modifies
the weight sampling step of standard ELM to improve performance as a surrogate
model for GSA. The method introduces a validation step to choose a sparsification parameter $p$. 
Similar to PCE, the Sobol' indices of SW-ELM, as defined in~\cite{ELM}, can be computed analytically. 
For the SW-ELM surrogates used in our experiments, the number of neurons used is half the number of training points.
A fraction of the training points are used as a validation set to choose the sparsification parameter. See~\cite{ELM} for further details.

\section{Method}\label{sec:method}
In this section, we outline our proposed approach for GSA of 
hyperparameter-to-statistic (HS) mappings of the form~\eqref{equ:hsmap}.
Our focus will be mainly on HS mappings that
involve integrating over the posterior. Examples are the posterior mean or
variance. For simplicity, we focus on 
\begin{equation}\label{equ:Fmean}
F(\boldsymbol\xi)=\Exi_\mathrm{post}(q(\boldsymbol\theta)) 
= \int_{\R^d}q(\boldsymbol\theta)  \pipostxi(\boldsymbol\theta)
\,d\vec\theta.
\end{equation}
It is straightforward to generalize the strategies described below t the cases of 
variance and higher order moments.
For brevity, we have suppressed the dependence of the 
posterior density on data $\vec{d}$ in~\cref{equ:Fmean}.

Computing the Sobol' indices of~\eqref{equ:Fmean} is often
challenging.  Computing $F(\vec\xi)$ via direct sampling requires generating 
samples from the posterior law of $\vec\theta$ using a Markov Chain Monte Carlo (MCMC)
method.  A naive approach for computing the Sobol' indices of $F(\vec\xi)$ would be
to follow a sampling procedure where an MCMC simulation is carried out for each
realization of $\vec\xi$.  This is typically infeasible.
For one thing, the computational cost of this naive approach will be prohibitive
for most practical problems.
In addition, performing multiple runs of an MCMC algorithm can be problematic, 
because such methods typically have algorithm-specific parameters that might 
need tuning for different realizations of $\vec\xi$. 

In~\cref{sec:HS_moment_IS}, we 
outline an approach that combines MCMC and importance sampling
for fast computation of moment-based HS mappings under study.
Then, in~\cref{sec:algorithm}, we present an algorithm that combines the
approach in~\cref{sec:HS_moment_IS} and surrogate models to 
facilitate GSA of moment-based HS maps. In that section, we also discuss 
the computational cost of the proposed approach, in terms of the 
number of required forward model evaluations. 
We also briefly discuss GSA of $F_\mathrm{MAP}$ in~\cref{sec:GSA_MAP}.

\subsection{Importance sampling for fast evaluation of moment-based 
HS maps}\label{sec:HS_moment_IS}
Importance sampling~\cite{TokdarKass10,OwenIS} aims at accelerating the computation of
integrals such as~\cref{equ:Fmean}, where the target distribution $\pipostxi$ is
difficult to sample from.  This is done by introducing an importance sampling
distribution $\pi_\mathrm{IS}(\boldsymbol\theta)$, which is tractable to work
with, and from which we are likely to sample points where the target posterior
distribution takes high density.  

Let $\pi_\mathrm{IS}$ be an importance sampling distribution.  The
integral~\cref{equ:Fmean}  can be written as 
\begin{equation}\label{equ:is}
\int_{\R^d} q(\boldsymbol\theta) \pipostxi(\vec\theta)\,d\vec\theta 
   = \int_{\R^d}
      \wxi(\vec\theta)q(\vec\theta)
		   \pi_\mathrm{IS}(\vec\theta)\,d\vec\theta,
		   \quad \text{with}\;
      \wxi(\vec\theta) = 
	  \frac{\pipostxi(\vec\theta)}
	       {\pi_\mathrm{IS}(\vec\theta)},
\end{equation}
provided that $\pi_\mathrm{IS}(\boldsymbol\theta)>0$ whenever $q(\boldsymbol\theta)\pi_\mathrm{post}(\boldsymbol\theta)\neq 0$~\cite{TokdarKass10}. When this holds, we can create a Monte Carlo estimate of~\eqref{equ:is}
\begin{equation}\label{equ:issum}
\int_{\R^d}q(\boldsymbol\theta)\pi_\mathrm{post}(\boldsymbol\theta;\boldsymbol\xi))d\boldsymbol\theta\approx \sum_{i=1}^M w_i \, q(\boldsymbol\theta_i) ,\quad \boldsymbol\theta_i\sim\pi_\mathrm{IS},
\end{equation}
where 
$w_i = \wxi(\vec\theta_i)$, $i = 1, \ldots, M$,
define the importance sampling weights. For our purposes, we desire weights that
are much greater than zero and have little variation over different samples. 
Our motivation for using importance sampling is to compute~\cref{equ:issum} for
different realizations of $\vec\xi$ without the need for multiple MCMC runs. 
Specifically, we propose an importance sampling approach tailored to the Bayesian inverse
problem of interest that enables
computing~\eqref{equ:issum} for different choices of  
$\vec\xi$ using the same importance sampling distribution.

\begin{figure}
\centering
\begin{tikzpicture}
\draw[densely dotted, ultra thick,rotate around={0:(0,0)}] (2.5,1.7) ellipse [x radius=2.2, y radius=1.5];
%
\draw[dashed,rotate around={30:(2.5, 1.7)}] (2.5, 1.7) ellipse [x radius=1.8, y radius=0.8];
\draw[solid,rotate around={-45:(2.5, 1.7) }] (2.5, 1.7)  ellipse [x radius=1.4, y radius=0.6];
%
\draw (-.3, 0) -- (5, 0) node[right] {};
\draw (0, -.3) -- (0, 3) node[above] {};
\node[below] at (4.95,1.295){$\pipris$};
\node[below] at (2.85,2.15){$\pi_\text{pr}^{\vec\xi_1}$};
\node[below] at (1.595,1.295){$\pi_\text{pr}^{\vec\xi_2}$};
\end{tikzpicture}
\caption{The interiors of the solid-line and dashed-line ellipses represent the
high density regions of two priors $\pi_\text{pr}^{\vec\xi_1}$ and
$\pi_\text{pr}^{\vec\xi_2}$, respectively. They are both enclosed by the high
density region of $\pipris$, represented by the interior of the dotted-line
ellipse.}\label{fig:isex}
\end{figure}
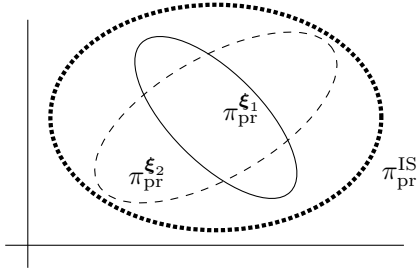

Since we consider choosing the prior distribution from a
parameterized family, the target posterior distributions
belong to a parameterized family (parameterized by the same prior
hyperparameters) as well. We let the importance sampling distribution 
be the posterior $\pi_\text{IS}=\pipostis$ constructed using a 
specific choice of prior, $\pipris$. 
This $\pipris$ is chosen from the same family as the priors in such a way 
that its high probability region covers that of the family of target priors. 
See~\cref{fig:isex} for an illustration, for the case of Gaussian priors. 
We then consider 
\begin{equation}\label{equ:pipostIS}
\pipostis(\vec \theta | \vec{d}) \propto
\pi_\mathrm{like}(\vec d | \vec \theta)\times\pipris(\vec \theta).
\end{equation} 
Importance sampling often breaks down if the
importance sampling distribution fails to cover the density of the target,
especially when the target distribution has a heavy tail.  As noted in our computational results, choosing
a prior that ``covers" all the target priors typically results in a
suitable importance sampling posterior $\pipostis$. 
 With the present strategy, it is possible to sample
from $\pipostis$ with one run of MCMC and gather information for all the target
posteriors. 

Next, we derive an expression for the estimator~\eqref{equ:issum} when
$\pi_\text{IS}=\pipostis$.  
We let $\vec\theta_\text{IS}$ and $\Gamma_\text{IS}$ denote the mean and
covariance of $\pipris$ while $\vec\theta_\vec{\xi}$ and $\Gamma_\vec{\xi}$ will
denote the mean and covariance of $\piprxi$. Let $\Pxi$ and $\Pis$ be the
normalization constants that correspond to $\pipostxi$ and $\pipostis$,
respectively:
\begin{equation}\label{equ:normcon}
\Pxi \defeq \int_{\R^d}\pi_\text{like}(\vec d|\vec\theta)\piprxi(\vec\theta)\,d\vec\theta,
\quad \Pis \defeq \int_{\R^d}\pi_\text{like}(\vec d|\vec\theta)\pipris(\vec\theta)\,d\vec\theta.
\end{equation}
We can write the importance sampling weights in~\eqref{equ:is} as
\begin{equation}
\wxi(\vec\theta)  = \frac{\pipostxi(\vec\theta)} {\pipostis\vec\theta)} 
	       =  \frac{\piprxi(\vec\theta)\pi_\text{like}(\vec\theta)/\Pxi}{\pipris(\vec\theta)\pi_\text{like}(\vec\theta)/\Pis}
	    =  \frac{1}{\Pxi/\Pis}\frac{\piprxi(\vec\theta)}{\pipris(\vec\theta)} \label{equ:isweight}.
\end{equation}
Letting the 
importance sampling weight in~\eqref{equ:issum} 
be given by~\cref{equ:isweight}, we obtain

\begin{equation}\label{equ:isestimator1}
\int_{\R^d} q(\boldsymbol\theta) \pipostxi(\vec\theta)\,d\vec\theta = 
\frac{1}{\Pxi/\Pis}\int_{\R^d}q(\vec\theta)\frac{\piprxi(\vec\theta)}{\pipris(\vec\theta)} \pipostis(\vec\theta)\,d\vec\theta.
\end{equation}
Furthermore, we can use the importance sampling distribution to rewrite 
the ratio of normalization constants $\Pxi/\Pis$ as
\begin{align}
\frac{\Pxi}{\Pis} = & \frac{1}{\Pis} \int_{\R^d}\pi_\text{like}(\vec d|\vec\theta)\piprxi(\vec\theta)\,d\vec\theta \notag\\
= & \frac{1}{\Pis} \int_{\R^d} \pi_\text{like}(\vec d|\vec\theta)\piprxi(\vec\theta) 
\frac{\Pis}{\pi_\text{like}(\vec d|\vec\theta)\pipris(\vec\theta)}  \pipostis(\vec\theta)\,d\vec\theta \notag\\
= & \frac{1}{\Pis} \int_{\R^d} \Pis \frac{\piprxi(\vec\theta) }{\pipris(\vec\theta)}  \pipostis(\vec\theta)\,d\vec\theta \notag\\
= & \int_{\R^d}  \frac{\piprxi(\vec\theta)}{\pipris(\vec\theta)}  \pipostis(\vec\theta)\,d\vec\theta\label{equ:normratio}.
\end{align}
Combining the expressions~\eqref{equ:isestimator1} and~\cref{equ:normratio} yields the estimator
\begin{equation}\label{equ:ismc}
F(\vec\xi) = \int_{\R^d} q(\boldsymbol\theta) \pipostxi(\vec\theta)\,d\vec\theta 
\approx\frac{1}{C(\vec\theta_1,\ldots,\vec\theta_M)}\sum_{i=1}^Mq(\boldsymbol\theta_i)\frac{\piprxi(\vec\theta_i)}{\pipris(\boldsymbol\theta_i)},\quad \vec\theta_i\sim\pipostis,
\end{equation}
where $C(\vec\theta_1,\ldots,\vec\theta_M) =
\sum_{i=1}^M\frac{\piprxi(\vec\theta_i)}{\pipris(\vec\theta_i)}$ is from the
estimator of~\cref{equ:normratio}. Note that in the case of Gaussian 
priors,
\begin{equation}
	\frac{\piprxi(\vec\theta)}{\pipris(\boldsymbol\theta)}=\mathrm{exp}\Big[\frac{1}{2}\left((\vec\theta_\mathrm{IS}-\boldsymbol\theta)^\top\Gamma_\mathrm{IS}^{-1}(\boldsymbol\theta_\mathrm{IS}-\boldsymbol\theta)-(\boldsymbol\theta_\vec{\xi}-\boldsymbol\theta)^\top\Gamma_\vec{\xi}^{-1}(\boldsymbol\theta_\vec{\xi}-\boldsymbol\theta)\right)\Big].
\end{equation}

There are some diagnostics for evaluating the effectiveness of a sample set from
the importance sampling distribution~\cite{OwenIS}. We use effective sample
size in our experiments.  A large effective sample size is desirable as it
indicates small variation in the estimator~\eqref{equ:isestimator1}.  
For a given $\vec\xi$, the effective sample size is
\begin{equation}\label{equ:effsamp}
n_\text{E}^\vec{\xi} \defeq \frac{\big(\sum_{i=1}^M w^\vec{\xi}(\vec\theta_i)\big)^2}
{\sum_{i=1}^M w^\vec{\xi}(\vec\theta_i)^2},\quad \vec\theta_i\sim\pipostis.
\end{equation}
Recall from~\eqref{equ:isweight} that we can rewrite $w^\vec{\xi} = \frac{\pipostxi}{\pipostis} = 
\frac{1}{P^\vec{\xi}/P^\text{IS}}\frac{\piprxi}{\pipris} $, 
Then, we can write~\eqref{equ:effsamp} as
\begin{equation}
	n_\text{E}^\vec{\xi} = 
\frac{\big(\sum_{i=1}^M \frac{\pipostxi(\vec\theta_i)}{\pipostis(\vec\theta_i)} \big)^2}{\sum_{i=1}^M
\big(\frac{\pipostxi(\vec\theta_i)}{\pipostis(\vec\theta_i)} \big)^2} 
= 
\frac{\big(\sum_{i=1}^M \frac{1}{P^\vec{\xi}/P^\text{IS}}\frac{\piprxi(\vec\theta_i)}{\pipris(\vec\theta_i)} \big)^2}{\sum_{i=1}^M
\big(\frac{1}{P^\vec{\xi}/P^\text{IS}}\frac{\piprxi(\vec\theta_i)}{\pipris(\vec\theta_i)} \big)^2} 
= 
\frac{\big(\sum_{i=1}^M \frac{\piprxi(\vec\theta_i)}{\pipris(\vec\theta_i)} \big)^2}{\sum_{i=1}^M
\big(\frac{\piprxi(\vec\theta_i)}{\pipris(\vec\theta_i)} \big)^2}.\label{equ:effsamp_pr}
\end{equation} 
In practice, we assess the suitability of $\pipostis$ as an importance sampling 
distribution by examining the distribution of $n_\text{E}^\vec{\xi}$ for an ensemble of 
realizations of $\xi$. This is illustrated in our computational results in 
Section~\ref{sec:numerics}.


\subsection{Algorithm for GSA of moment based HS maps}\label{sec:algorithm}
We approximate $F_\mathrm{mean}(\boldsymbol\xi)$ and
$F_\mathrm{var}(\boldsymbol\xi)$ using~\eqref{equ:ismc}.  When estimating their
Sobol' indices, we opt for the surrogate-assisted approach.
Because~\eqref{equ:ismc} provides us with noisy evaluations of
$F_\mathrm{mean}(\boldsymbol\xi)$ and $F_\mathrm{var}(\boldsymbol\xi)$,
estimating Sobol' indices by the double-loop sampling approach can give poor
results. Regression-based surrogate methods are well-suited here. We employ
sparse regression PCE and sparse-weight ELM, discussed in~\cref{sec:gsa}.
In~\cref{alg:spelm}, we detail the proposed framework for variance-based GSA
of~\eqref{equ:hsmap}. Samples from one MCMC run are used to estimate, by
importance sampling, $F(\boldsymbol\xi)$ for selected realizations of
$\boldsymbol\xi$. 
Sample realizations of $\vec\xi$ are generated by Latin hypercube sampling (LHS)~\cite{LHS79,Viana16}. . 
 These samples serve as a training set for building surrogate
models of $F(\boldsymbol\xi)$ for GSA, as discussed in~\cref{sec:gsa}. 
The purpose of using two different surrogate methods is to help 
gain further confidence in the computed results.

\begin{algorithm}[h]
\caption{GSA with respect to uncertain prior hyperparameters via importance sampling}\label{alg:spelm}
\textbf{Input:} (i) Likelihood PDF $\pi_\mathrm{like}(\vec d|\vec\theta)$
(ii) Hyperparameter-dependent prior PDF $\piprxi(\boldsymbol\theta)$
(iii) Importance sampling prior PDF $\pipris(\boldsymbol\theta)$
(iv) Collection of hyperparameter samples $\{\boldsymbol\xi_k\}_{k=1}^N$ 
(v) QoI function $q(\boldsymbol\theta)$
(vi) Monte Carlo sample size $M$

\textbf{Output:} 
(i) First-order Sobol' indices
(ii) Total Sobol' indices
\begin{algorithmic}[1]
\State Perform MCMC to generate samples, $\{\boldsymbol\theta_i\}_{i=1}^M$, from $\pipostis$, of which $\widehat M$ are distinct
\For{$i=1,\dots,\widehat M$}
	\State Compute and store $q(\boldsymbol\theta_i)$
	\State Compute and store $\pipris(\boldsymbol\theta_i)$
	\For{$k=1,\dots,N$}
	\State Compute and store $\pi_\text{pr}^{\vec\xi_k}(\boldsymbol\theta_i)$
	\EndFor
\EndFor
\For{$k=1,\dots,N$}
\State Approximate $F(\boldsymbol\xi_k)$ in~\eqref{equ:hsmap} using the estimator~\eqref{equ:ismc} with $\{(q(\boldsymbol\theta_i),\pi_\text{pr}^{\vec\xi_k}(\boldsymbol\theta_i),\pipris(\boldsymbol\theta_i)\}_{i=1}^M$
\EndFor
\State Compute a surrogate model $\widehat{F}$ for $F$, using 
the samples $\{(\boldsymbol\xi_k,F(\boldsymbol\xi_k)\}_{k=1}^N$
\State Estimate first-order and total Sobol' indices of $\widehat{F}$ 
\end{algorithmic}
\end{algorithm}

Under the assumption that the model and QoI $q$ are expensive to
evaluate,~\cref{alg:spelm} incurs most of its cost during the MCMC sampling
stage. In this work, we use the delayed-rejection adaptive Metropolis
(DRAM)~\cite{Haario06,Laine08,Smith14} algorithm to perform MCMC. With
delayed-rejection, each MCMC stage can include up to a fixed number of extra
delayed-rejection steps. Each of these steps requires us to evaluate the model
an additional time. Typically, one initially runs MCMC for $M_\text{burn}$
burn-in stages. These burn-in samples are discarded and not included in the set
of posterior draws. The cost of running the MCMC stage in~\cref{alg:spelm} with DRAM
is $\mathcal{O}(M+M_\text{burn})$ model evaluations. 
In the second stage, we evaluate $q$ at the distinct MCMC samples. 
Because the MCMC samples usually include repeated draws, the number of these QoI 
evaluations is less than $M$.

\subsection{GSA of the MAP point}\label{sec:GSA_MAP}
The MAP point is an important point estimator and studying its sensitivity to
prior parameters complements the study of other moment-based HS maps such as the
posterior mean or variance.  The approach described in~\cref{alg:spelm} can be
used in cases where $F(\boldsymbol\xi)$ involves moments of the posterior, as in
the case of the mean and variance.  On the other hand, evaluating
$F_\mathrm{MAP}$ requires solving the regularized nonlinear least squares
problem~\eqref{equ:mappoint} for each realization of $\vec\xi$. No numerical
integration is needed.  One does not even need to know the
normalization constant of the posterior to find its MAP point. While we do not
use~\cref{alg:spelm} to study $F_\mathrm{MAP}$, we evaluate it at the same set
of realizations $\{\vec\xi_k\}_{k=1}^N$ used in~\cref{alg:spelm}. These
evaluations are used to build surrogate models for $F_\text{MAP}$. The computed
surrogate is then used for fast GSA of $F_\text{MAP}$.

\section{Computational results}\label{sec:numerics}
In this section, we consider two model inverse problems as testbeds for our
proposed approach.  Specifically, we use~\cref{alg:spelm} for global sensitivity
analysis (GSA) of hyperparameter-to-statistic (HS) mappings from the inverse
problems under study. These examples are used to examine various aspects of the
proposed method.
In~\cref{sec:linear}, we consider a simple linear inverse problem. 
Specifically, we formulate fitting a line to noisy data as a linear
Bayesian inverse problem.  In this case, the posterior distribution is known
analytically.  This means that the HS mappings admit analytical forms, and we can
perform GSA without~\cref{alg:spelm}.  This problem serves as a
benchmark where we gauge the accuracy of GSA with~\cref{alg:spelm} against
reference values.  The QoI in this example is a quadratic function.
For this QoIs, we study the HS mappings for the mean and
variance.  The Sobol' indices, approximated using~\cref{alg:spelm}, of
these HS mappings are compared to the true Sobol' indices.  Overall, 
we note close agreement between the results produced by our method and the analytic
results.

Next, we apply our method to a nonlinear Bayesian inverse problem
in~\cref{sec:seir}.  The inverse problem is governed by an SEIR model from
epidemiology~\cite{SEIR00,ModernEp}.  It exemplifies the type of problem that~\cref{alg:spelm} is
designed and intended for.  Our numerical results provide a unique perspective
on the impact of uncertainty in prior hyperparameters.  The QoI is the basic
reproductive number.  We quantify the uncertainty in the mean, variance, and MAP
point that is caused by uncertainty in the prior hyperparameters.  The Sobol'
indices of the mean, variance, and MAP point HS mappings are computed
using~\cref{alg:spelm} and highlight the most influential hyperparameters in
each case.  We use two different surrogate modeling approaches in these
computations: one based on sparse polynomial chaos expansions (PCEs) and the
other based on sparse weight extreme learning machines (SW-ELMs).  The two
approaches provide results that match closely. 

\subsection{Linear Bayesian inverse problem}\label{sec:linear}

We consider the problem  of fitting a line $y=mt+b$ to noisy
measurements $\{(t_i,y_i)\}_{i=1}^{4}$ at times $t = 0,0.5,1.5,2.5$.  The slope
$m$ and intercept $b$ are treated as unknown parameters, which we seek to
estimate.  We cast this problem in a Bayesian framework. 
This serve to illustrate various properties of our proposed 
framework.

\subsubsection{Bayesian inverse problem setup}
Let the inversion parameter vector be denoted by 
$\boldsymbol{\theta}  = \big[\begin{array}{cc}
    b & m
\end{array}\big]^\top$.
We consider estimation of $\boldsymbol{\theta}$ from 
\begin{equation}\label{equ:lineariv}
\mat A \boldsymbol{\theta} + \boldsymbol{\eta} = \boldsymbol{y},
\end{equation}
where
Here $\mat A = \left[\begin{array}{cccc}
1 & 1 & 1 & 1\\
0 & 0.5 & 1.5 & 2.5
\end{array}\right]^\top$ is the forward operator,
$\vec{\eta}$ models measurement noise, and $\vec{y}$ is the data. 

\begin{figure}[h!!]
    \centering
   \subfigure{\includegraphics[width=0.45\textwidth]{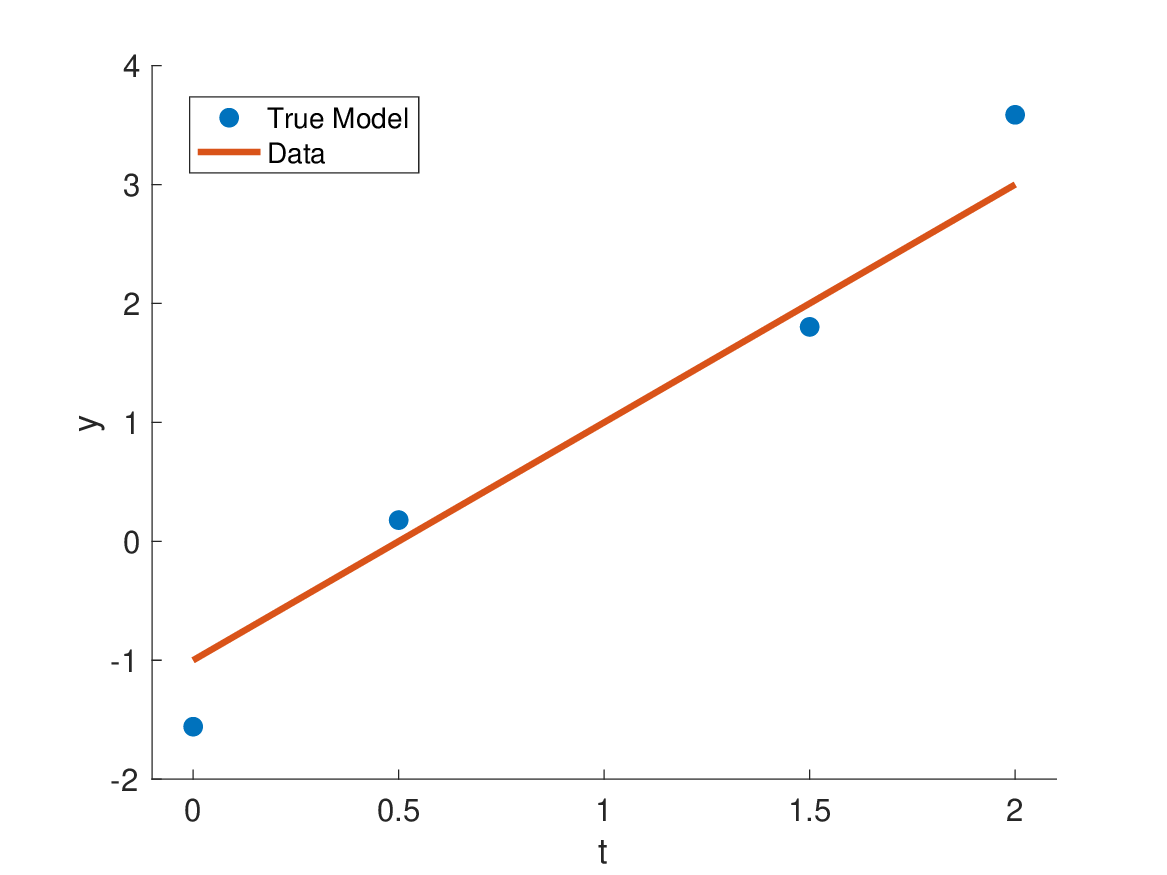}}
    \caption{The true trajectory of the linear model plotted with the noisy measurements at times $t=0,0.5,1.5,2$.}
    \label{fig:lin_traj}
\end{figure}

We assume noise at each measurement independently follows the standard normal
distribution, i.e.,  $\eta_i\sim \mathcal{N}(0,1)$. The noise covariance is
$\boldsymbol{\Gamma}_\mathrm{noise}=\mat I_{4\times 4}$.  
We assume a ``ground-truth'' parameter 
vector $\boldsymbol\theta_\text{true} = 
[1 \; -2]^\top$ and 
generate
measurements by adding sampled noise $\eta_i$ to $y_i = -2t_i + 1$ for
$i=1, \ldots, 4$; see~\cref{fig:lin_traj}.
We assume a Gaussian prior distribution $\mathcal{N}(\vec\theta_\text{pr},\vec\Gamma_\text{pr})$ 
for the inversion parameters $\vec\theta$ with
\begin{equation}\label{equ:lin_pr}\boldsymbol\theta_\mathrm{pr}=  \left[\begin{array}{cc}
 \mu_b\\
    \mu_m\end{array} \right],\quad  \boldsymbol\Gamma_\mathrm{pr}=\left[\begin{array}{cc}
\sigma_b^2 & 0 \\
0 & \sigma_m^2
\end{array}\right].
\end{equation}
Due to linearity of the parameter-to-observable map and Gaussian 
prior and noise models, 
the posterior distribution for $\vec\theta$ is also Gaussian and 
explicitly known.
It is the Gaussian distribution $\mathcal{N}(\vec\theta_\text{post},\vec\Gamma_\text{post})$, where
\begin{equation}\label{equ:lin_post}
\boldsymbol{\Gamma}_\mathrm{post}=(\mat A^\top\boldsymbol{\Gamma}_\mathrm{noise}^{-1}\mat A + \boldsymbol{\Gamma}_\mathrm{pr}^{-1})^{-1},\quad
\boldsymbol{\theta}_\mathrm{post}=\boldsymbol{\Gamma}_\mathrm{post}(\mat 
 A^\top \boldsymbol{\Gamma}_\mathrm{noise}^{-1}\boldsymbol{y}+\boldsymbol{\Gamma}_\mathrm{pr}^{-1}\boldsymbol{\theta}_\mathrm{pr}). 
\end{equation}
Since the posterior distribution is Gaussian, the posterior mean
and MAP point are the same.

\textbf{Quantity of interest.} We introduce the QoI
which depends on the inversion parameters $\vec\theta$. The QoI is the quadratic form
\begin{equation}\label{equ:lin_qoi2}
q(\vec\theta) = \vec\theta^\top\vec\theta = m^2 + b^2,\quad 
\vec\theta\sim\mathcal{N}(\vec\theta_\text{post},\vec\Gamma_\text{post}).
\end{equation}
 As $\vec\theta$ is a Gaussian random variable, 
we have access to expressions for the first and second moments~\cite{MatrixCookbook,GaussianMoments}. 
of the QoI. 
We can therefore express the mean and variance of the QoI analytically.
\begin{equation}\label{equ:meanvar_qoi2}
\mathbb{E}_\text{post}(q) = \vec\theta_\text{post}^\top\vec\theta_\text{post},
\quad \text{var}(q) = 2\ \text{tr}(\vec\Gamma_\text{post}^2) + 
4\vec\theta_\text{post}^\top\Gamma_\text{post}\vec\theta_\text{post}.
\end{equation} 
\textbf{Uncertainty in prior hyperparameters.}
Before building the posterior distribution, we must choose values for the prior
hyperparameters are $\vec\xi =\big[\begin{array}{cccc} \mu_b & \mu_m &
\sigma_b^2 & \sigma_m^2 \end{array}\big]^\top$ that appear
in~\eqref{equ:lin_pr}. 
We assume these parameters are specified within some interval around their
nominal values and are modeled as independent uniformly distributed random
variables.  We use a nominal value of 1 for each of the
parameters and let the upper and lower bounds of the distributions be $\pm
50\%$ perturbations of the nominal value.
 
 \subsubsection{Parameter estimation and importance sampling}
\begin{figure}[h!!]
    \centering
   \subfigure{\includegraphics[width=0.44\textwidth]{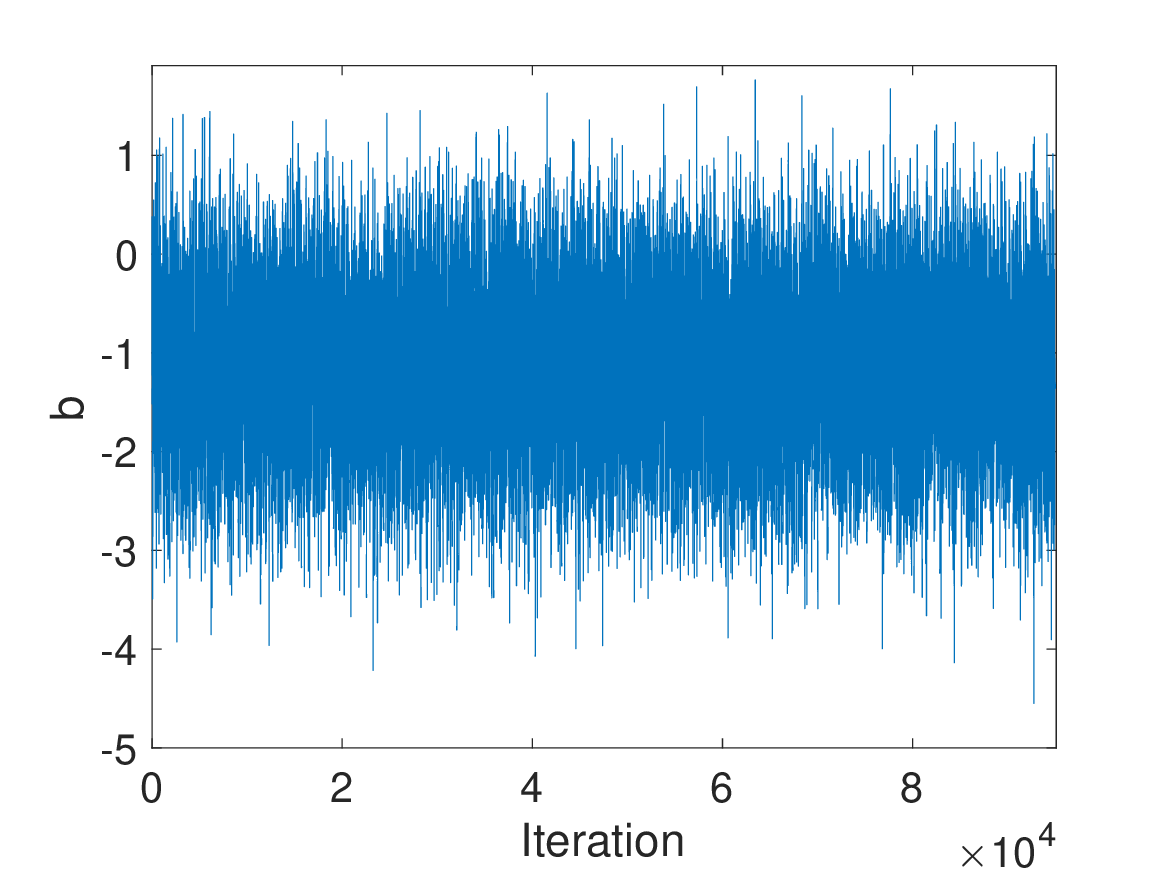}}
    \subfigure{\includegraphics[width=0.44\textwidth]{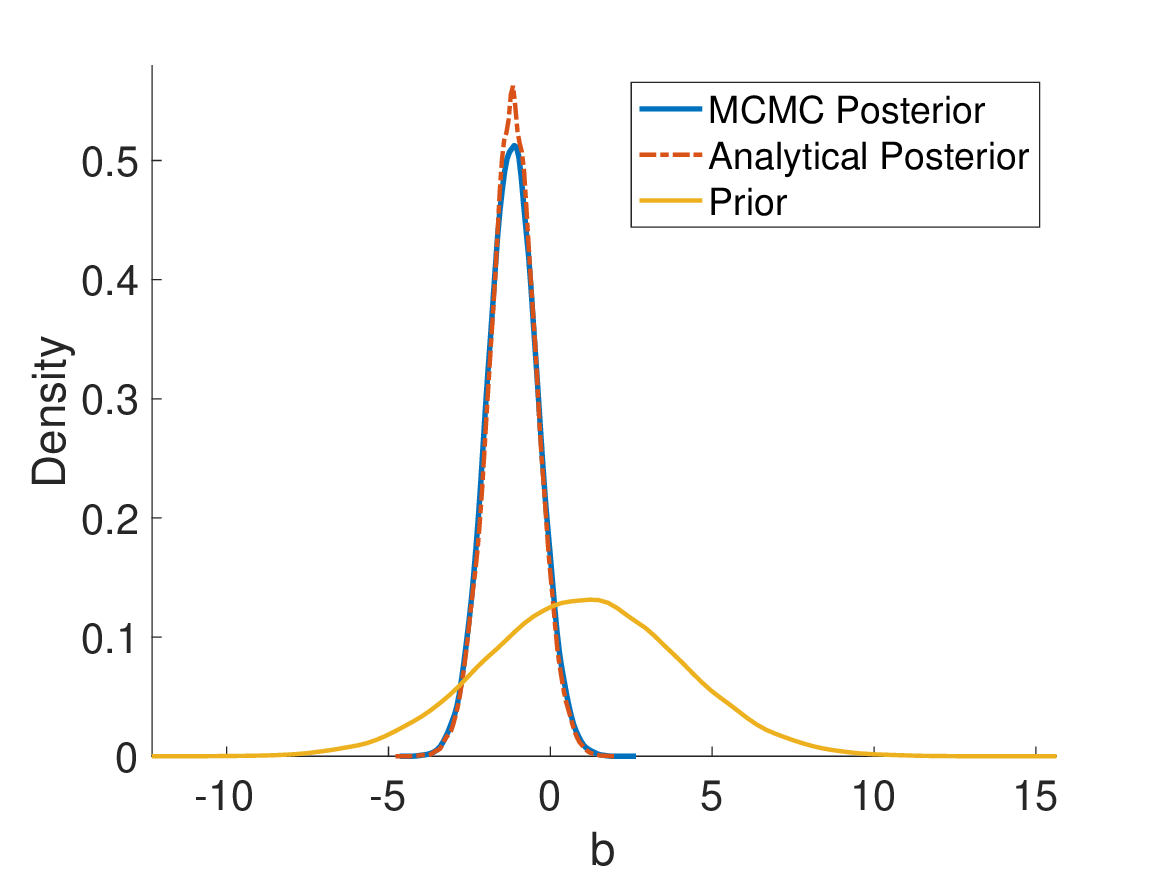}} 
       \subfigure{\includegraphics[width=0.44\textwidth]{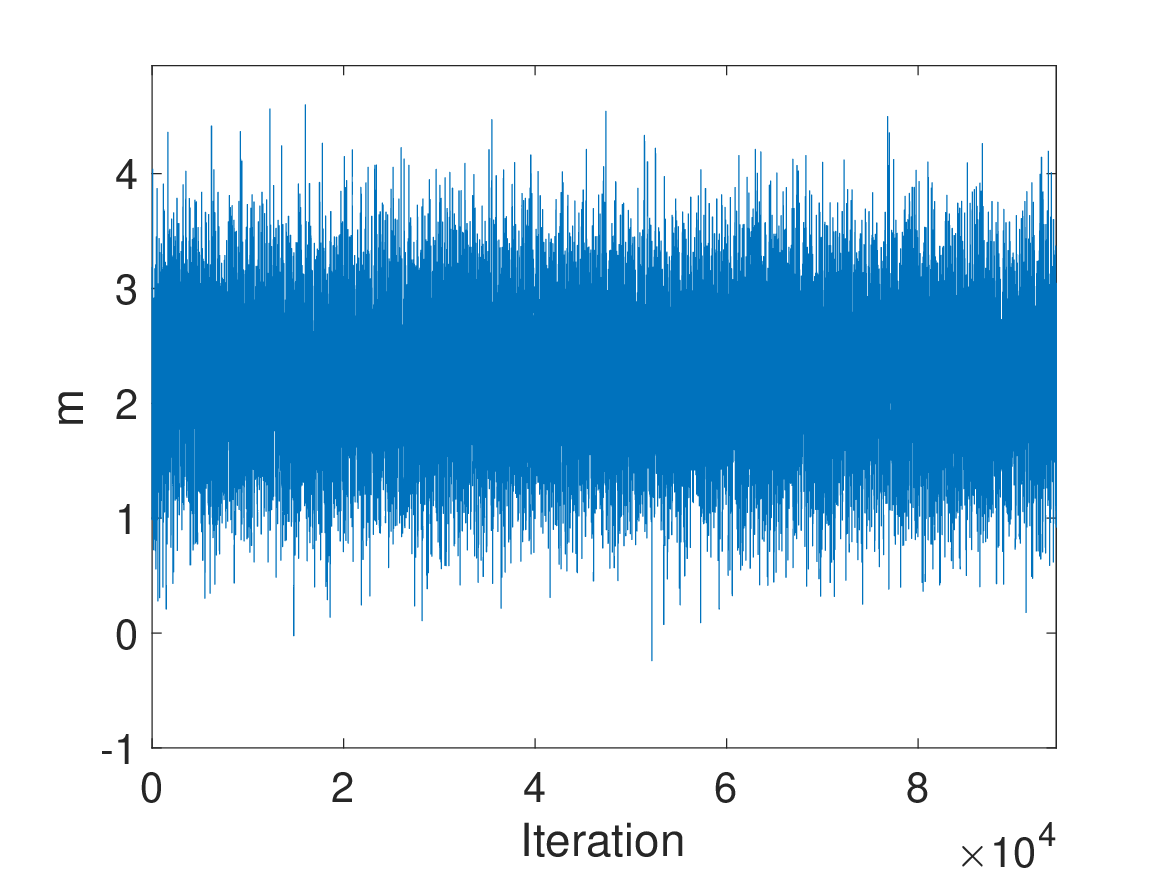}}
    \subfigure{\includegraphics[width=0.44\textwidth]{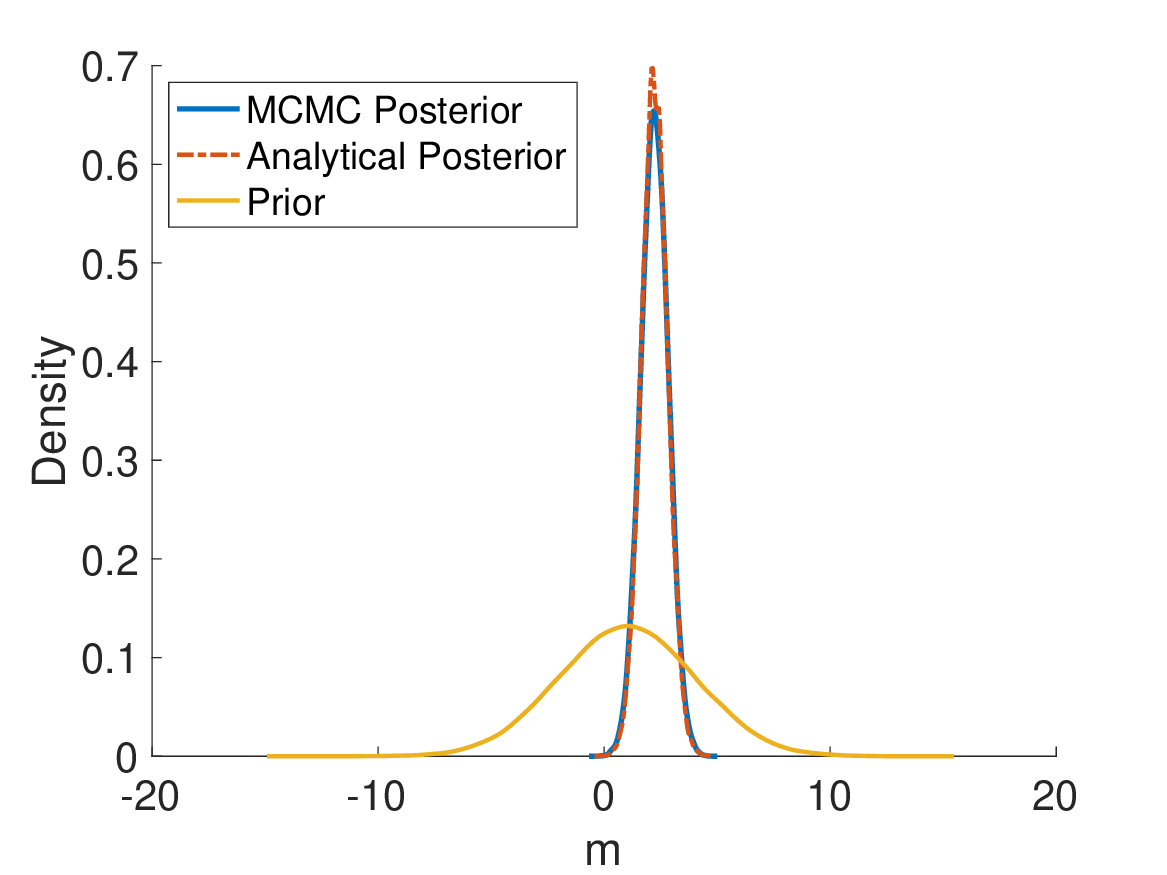}} 
    \caption{MCMC chains of the inversion parameters $m$ and $b$ along with corresponding marginal posterior distributions compared to prior distributions.}
    \label{fig:chain}
\end{figure}

\begin{figure}[h!!]
    \centering
   \subfigure{\includegraphics[width=0.44\textwidth]{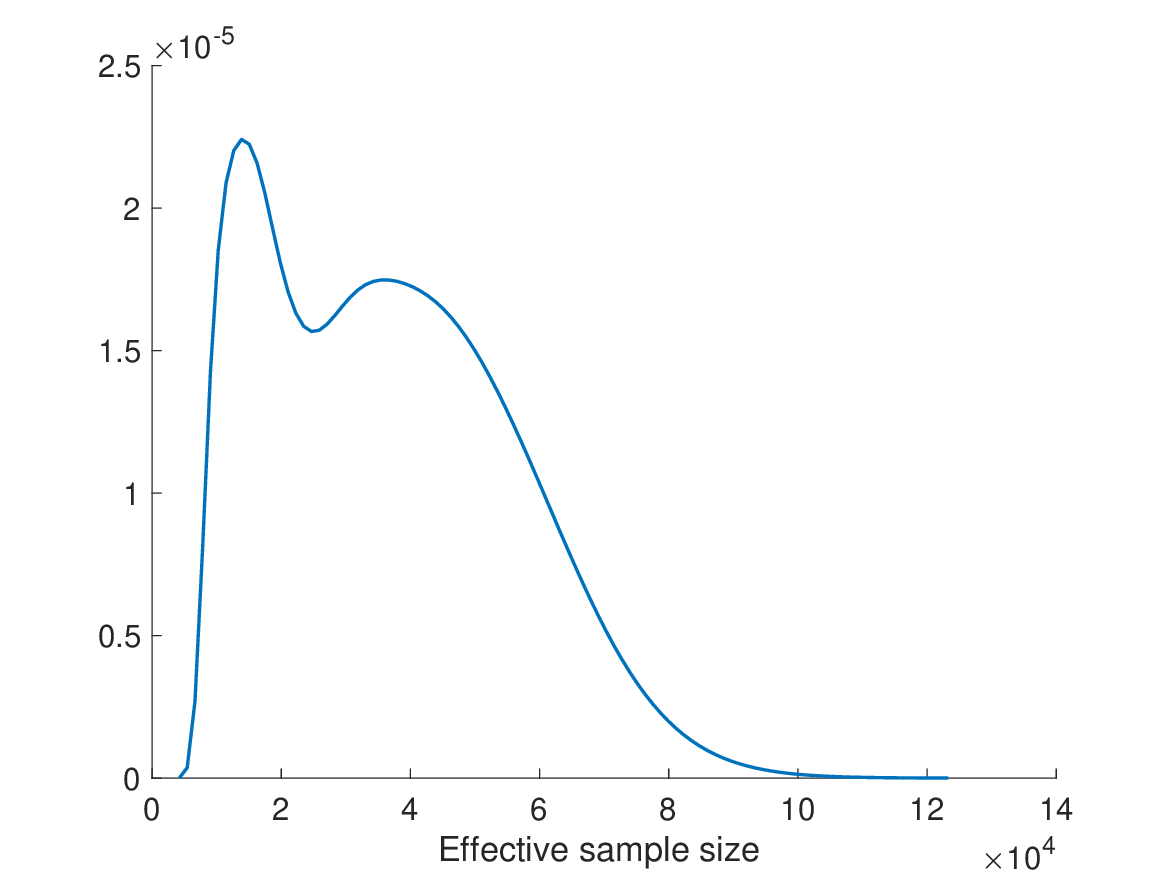}}
          \subfigure{\includegraphics[width=0.44\textwidth]{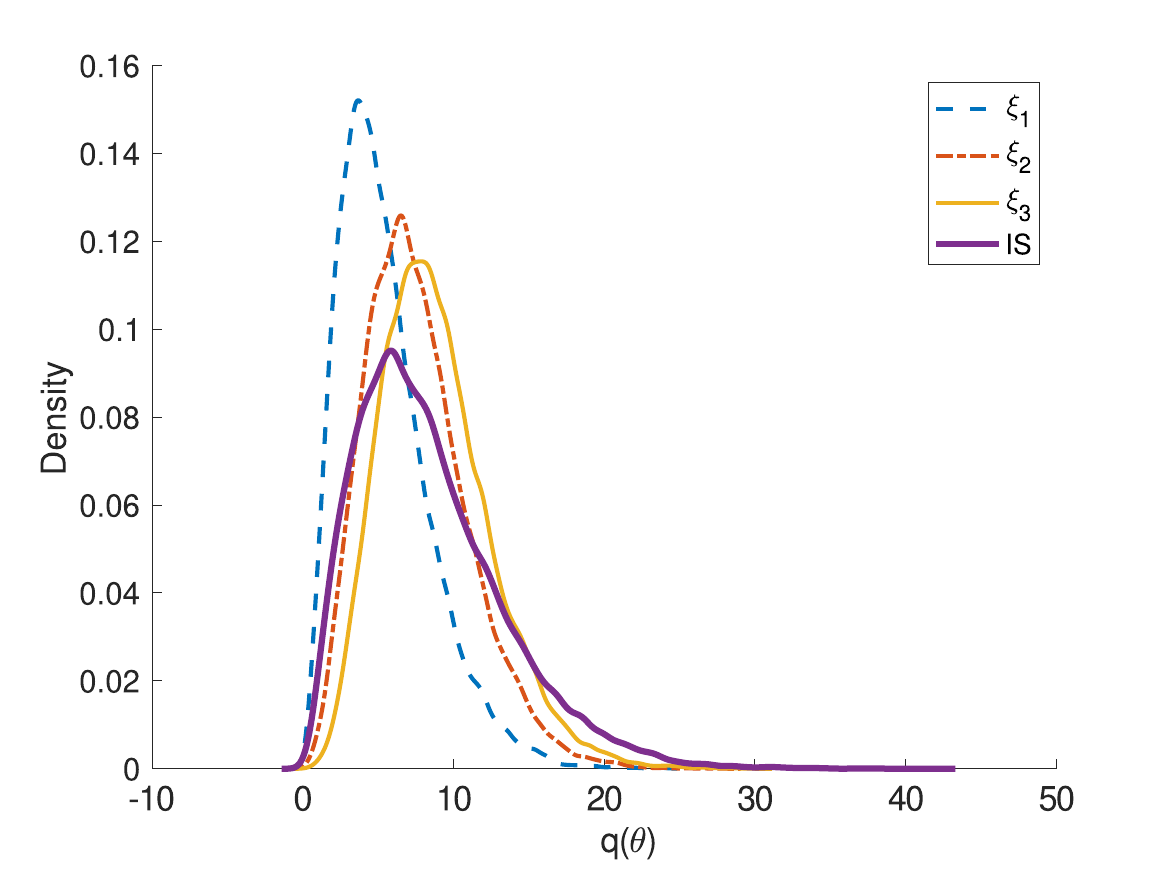}}  
    \caption{Left: Effective sample size of $\pipostis$ distributed over values $\vec\xi$.
     Right: Distribution of $q$ when $\vec\theta\sim\pipostis$ compared 
     to when $\vec\theta\sim\pipostxi$ for three realizations of $\vec\xi$. }
    \label{fig:lin_diag}
\end{figure}

 To understand how the uncertainty in the prior hyperparameters affects the QoI, we employ~\cref{alg:spelm} in~\cref{sec:method}. 
 The first step is to choose a prior $\pipris$ to build the importance sampling distribution $\pipostis$. We take $\mathcal{N}(\vec\theta_\text{pr}^\text{IS},\vec\Gamma_\text{pr}^\text{IS})$ with
 \begin{equation}\label{equ:lin_pr_IS}\boldsymbol\theta_\mathrm{pr}=  \left[\begin{array}{cc}
 1\\
    1\end{array} \right],\quad  \boldsymbol\Gamma_\mathrm{pr}=\left[\begin{array}{cc}
1.5^2 & 0 \\
0 & 1.5^2
\end{array}\right].
\end{equation}
We use the DRAM algorithm, discussed
 in~\cref{sec:method}, to draw $10^5$ samples from $\pipostis$. 
 In~\cref{fig:chain}, we compare the prior, analytic posterior, and MCMC-constructed posterior marginal distributions of $b$ and $m$.
Before we implement~\cref{alg:spelm}, we evaluate whether 
$\pipostis$ is an acceptable importance sampling distribution. 
As discussed in~\cref{sec:method}, we use~\eqref{equ:effsamp_pr} to compute 
the effective sample size over the distribution of prior hyperparameters $\vec\xi$. 
The distribution of effective sample sizes, given in~\cref{fig:lin_diag}~(left), shows 
that $\pipostis$ is an effective importance sampling distribution over many realizations of $\vec\xi$. 
In~\cref{fig:lin_diag}, we give a further visual of how $\pipostis$ 
serves as an effective importance sampling distribution. 
In the right panel, the distribution of $q$, 
when $\vec\theta\sim\pipostis$, is compared to the distributions of 
$q_\text{lin}(\vec\theta)$ when $\vec\theta\sim\pipostxi$, for three realizations 
of $\vec\xi$. 

\subsubsection{Sensitivity analysis}
We now study $q$ given in~\eqref{equ:lin_qoi2}.  We are
interested in the variance and mean HS mappings ~\eqref{equ:hsmap}
$F_\text{mean}(\vec\xi) =  \mathbb{E}_\text{post}^\vec{\xi}(q)$ and
$F_\text{var}(\vec\xi) = \text{var}^\vec{\xi}(q)$.  As shown
in~\cref{equ:meanvar_qoi2}, these HS mappings take analytically known forms.

We use~\cref{alg:spelm} to compute the Sobol' indices of the HS mappings under
study. The importance sampling distribution is given by $\pipostis$, as
described in \cref{sec:method}, and with $\pipris$ as specified
in~\cref{equ:lin_pr_IS}. We study how the Sobol' indices, computed
via~\cref{alg:spelm} converge as we increase MCMC sample size $M$.
In our computations, we build sparse PCE and sparse-weight ELM~\cite{ELM}
surrogate models, discussed in~\cref{sec:gsa} using $10^3$ realizations of
$\vec\xi$, drawn using Latin hypercube sampling (LHS).  SW-ELM surrogates use
$800$ realizations for training and $200$ for validation during the weight
sparsification step.  The Sobol' indices estimated by~\cref{alg:spelm} are
compared against benchmark indices.  We compute the benchmark indices by
applying the standard sampling approach from~\cite{Saltelli10} to the HS
mappings.  This yields accurate indices because we have
access to the analytic expressions of $F_\mathrm{var}$ and $F_\mathrm{mean}$.

\begin{figure}[h!!]
    \centering
   \subfigure[PCE]{\includegraphics[width=0.45\textwidth]{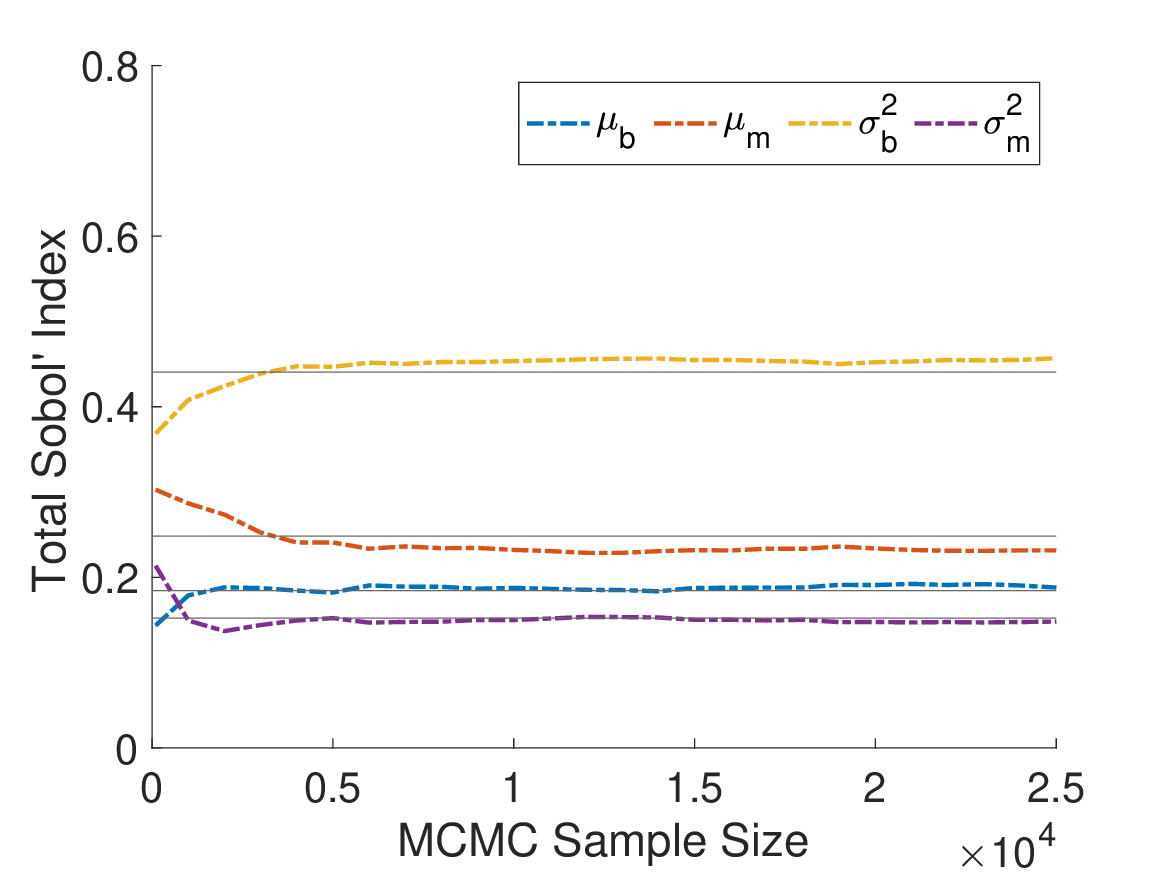}}
\subfigure[SW-ELM]{\includegraphics[width=0.45\textwidth]{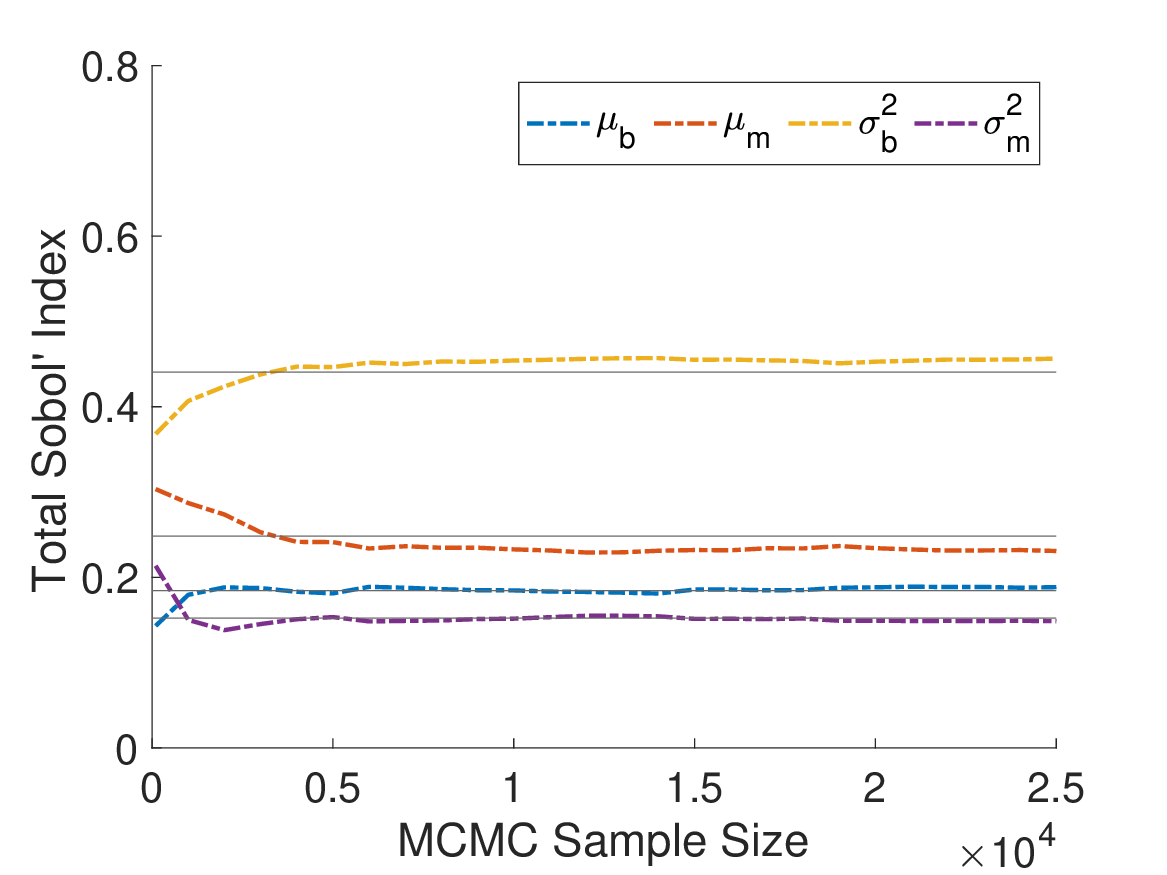}}
\subfigure[First-order Sobol' indices of $F_\text{mean}$]{\includegraphics[width=0.45\textwidth]{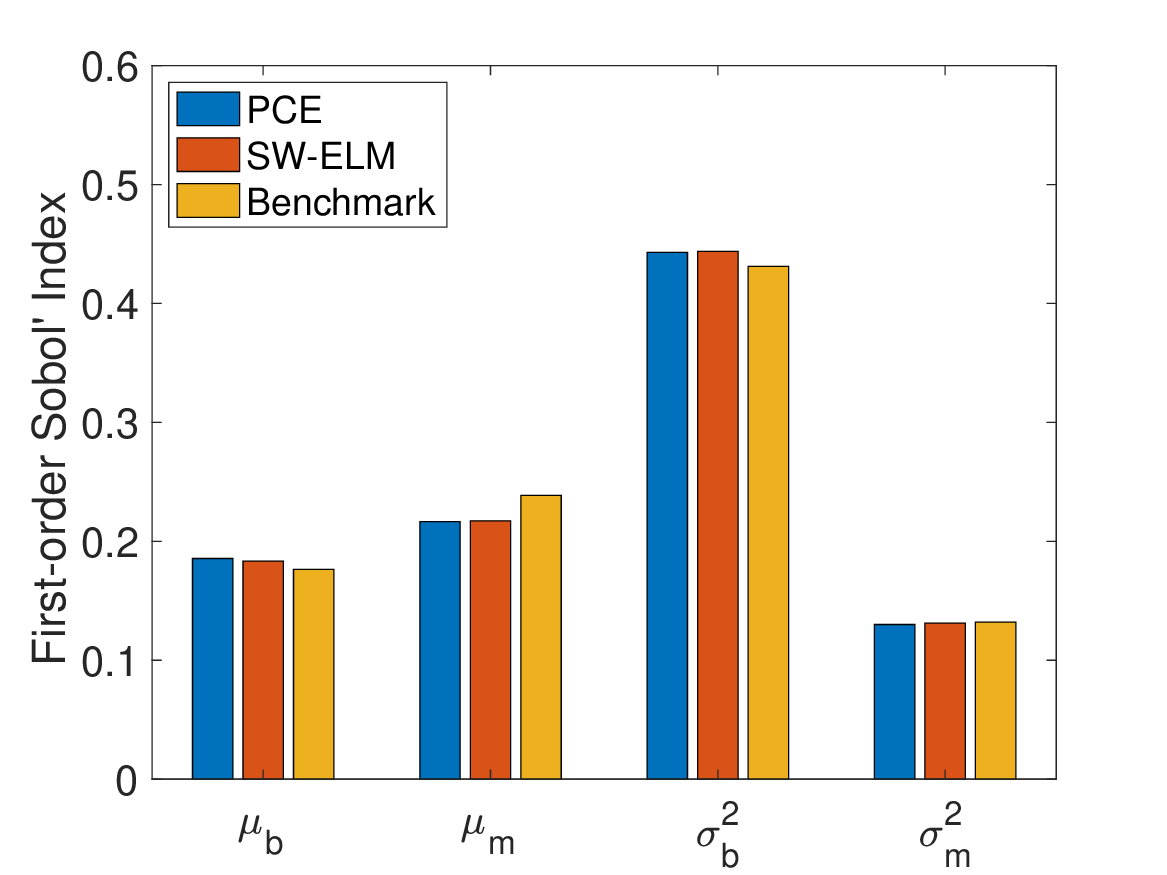}}
\subfigure[Total Sobol' indices of $F_\text{mean}$]{\includegraphics[width=0.45\textwidth]{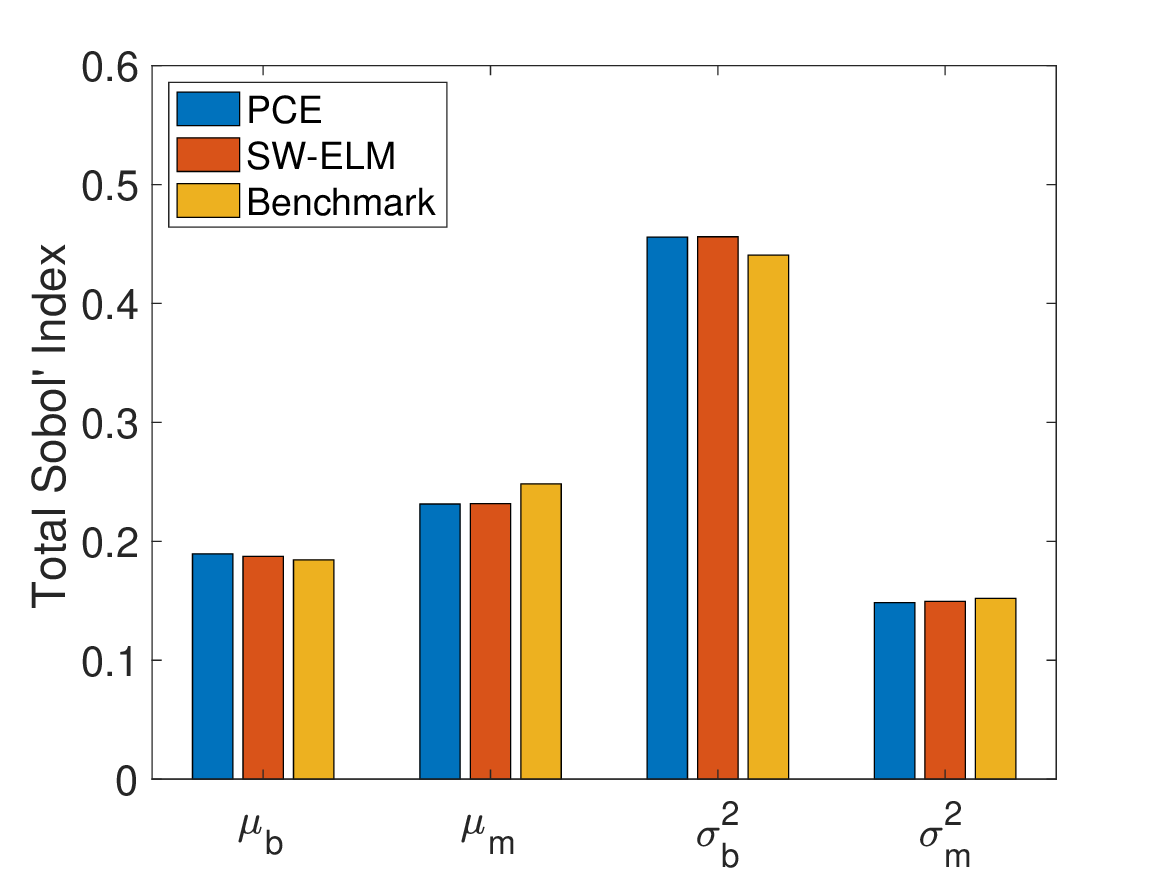}}
    \caption{Top left: Convergence experiment for $F_\text{mean}$ using PCE surrogate. 
    Top right: Convergence experiment for $F_\text{mean}$ using SW-ELM surrogate.  
    Bottom left: Comparison of true first-order Sobol' indices and surrogate-estimated indices 
    where $F_\text{mean}$ evaluations approximated with $M=9^4$ MCMC samples.  Bottom right: Comparison of true total Sobol' indices and surrogate-estimated indices where $F_\text{mean}$ evaluations approximated with $M=9^4$ MCMC samples.}
\label{fig:munon_sob}
\end{figure}

~\cref{fig:munon_sob} illustrates the Sobol' indices of $F_\text{mean}$.  The
computed indices are compared against benchmark values which are computed by
sampling the analytic form of the QoI.
Note that With only a modest number of about 1000 MCMC samples, we can ascertain
the correct importance ranking of the total Sobol' indices of $F_\text{mean}$.
By 5000 samples, the total Sobol' indices have converged. Also, 
as before, the two surrogate modeling 
approaches provide similar results. 

\begin{figure}[h!!]
   \centering
   \subfigure[PCE]{\includegraphics[width=0.45\textwidth]{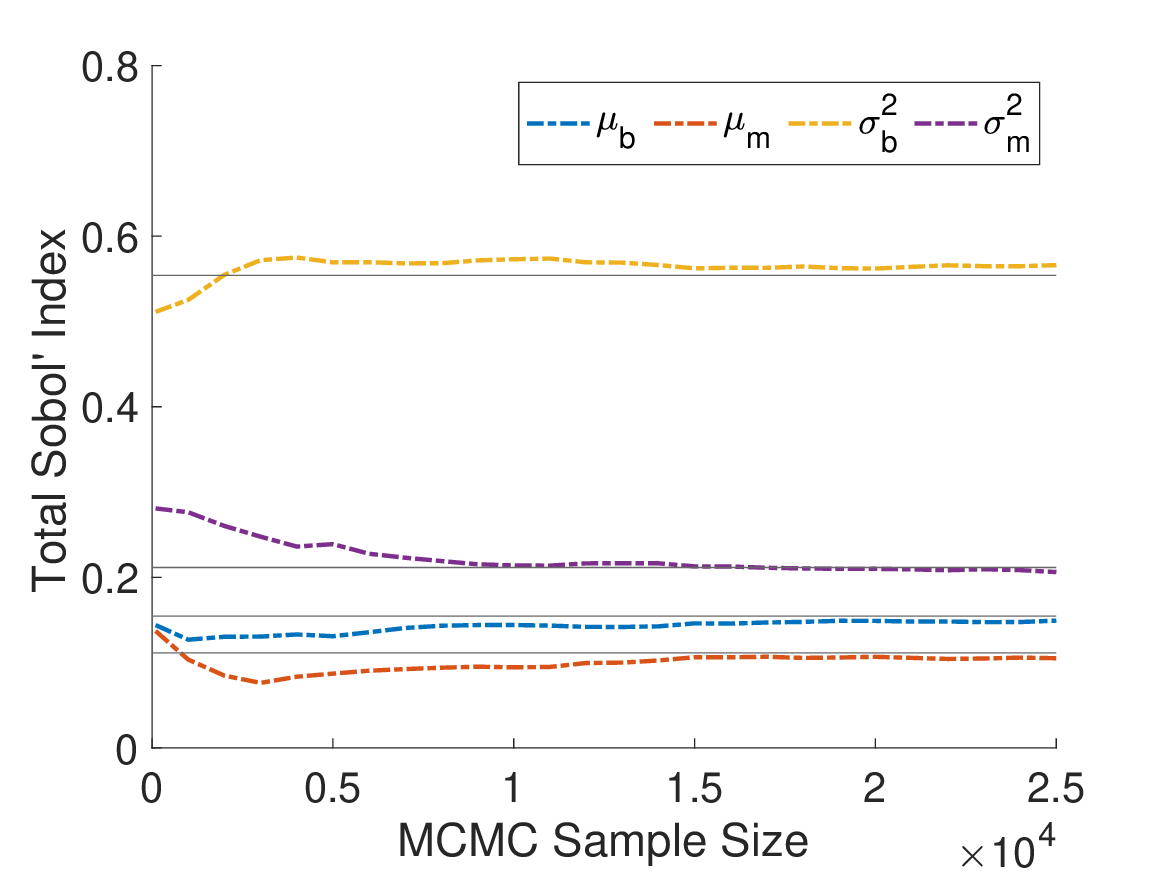}}
\subfigure[SW-ELM]{\includegraphics[width=0.45\textwidth]{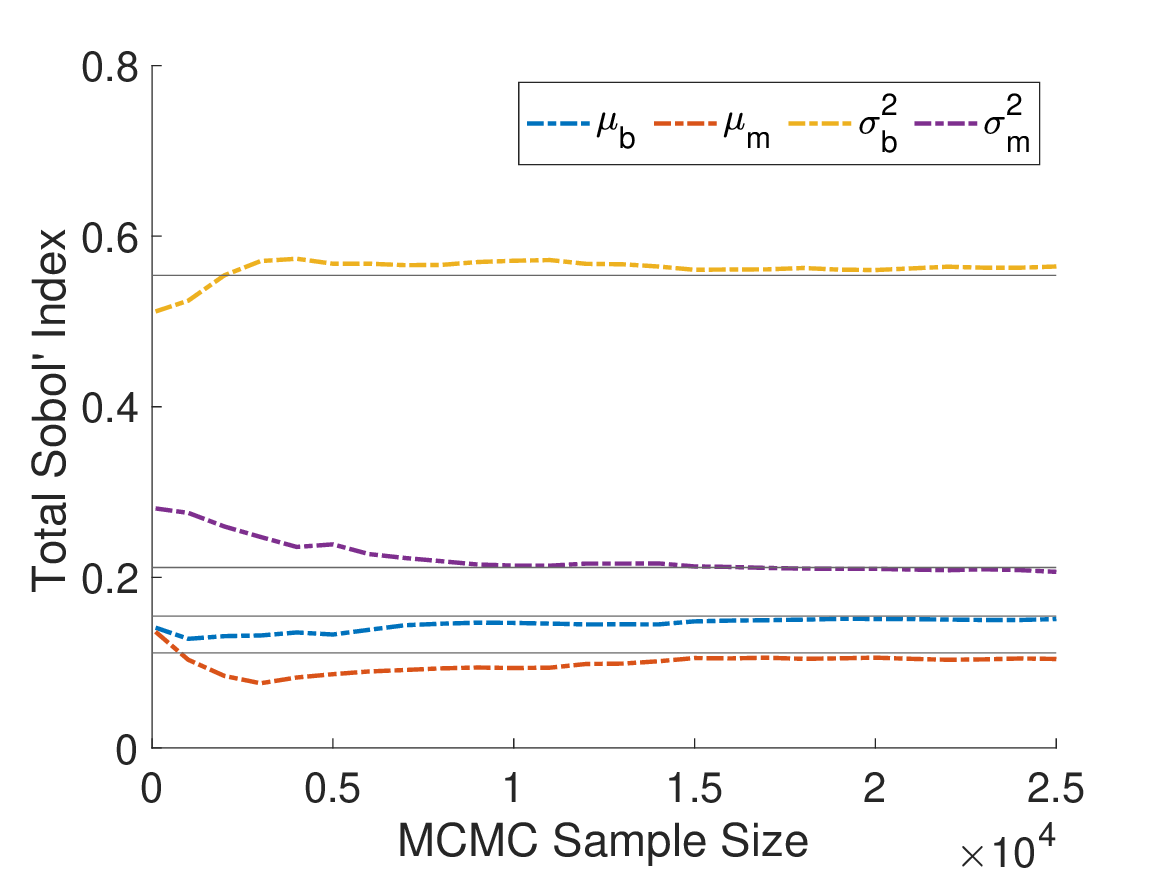}}
\subfigure[First-order Sobol' indices of $F_\text{var}$]{\includegraphics[width=0.45\textwidth]{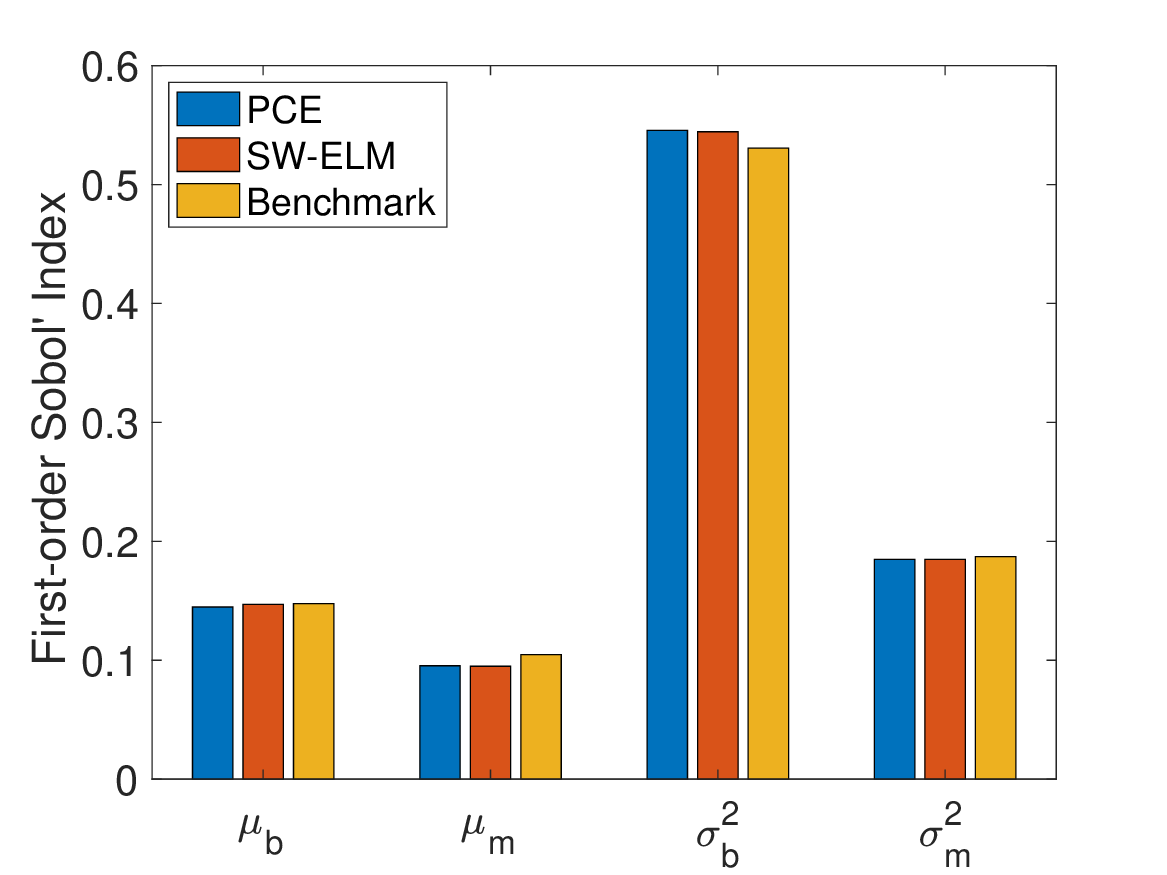}}
\subfigure[Total Sobol' indices of $F_\text{var}$]{\includegraphics[width=0.45\textwidth]{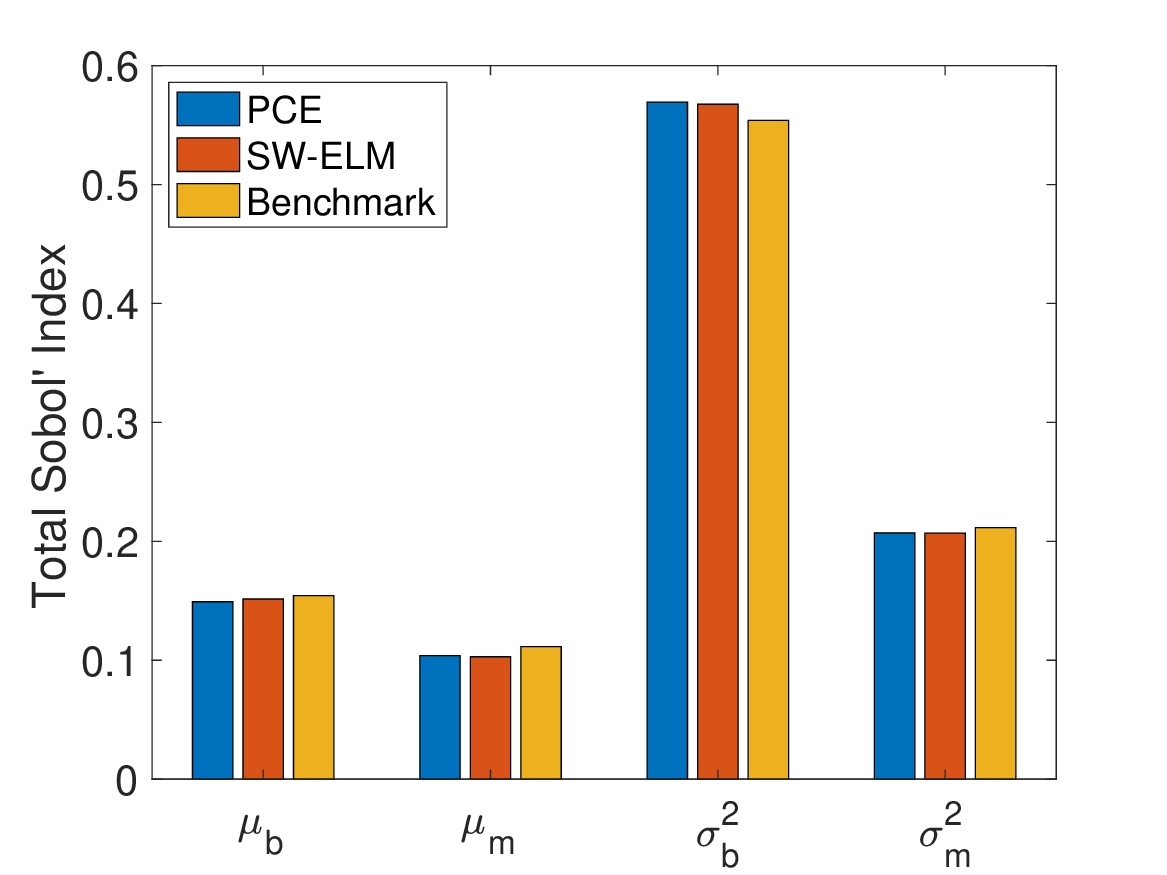}}
    \caption{Top left: Convergence experiment for $F_\text{var}$ using PCE surrogate. 
    Top right: Convergence experiment for $F_\text{var}$ using SW-ELM surrogate.  
    Bottom left: Comparison of true first-order Sobol' indices and surrogate-estimated indices 
    where $F_\text{var}$ evaluations approximated with $M=9^4$ MCMC samples.  Bottom right: Comparison of true total Sobol' indices and surrogate-estimated indices where $F_\text{var}$ evaluations approximated with $M=9^4$ MCMC samples.}
    \label{fig:vanon_sob}
\end{figure}

In~\cref{fig:vanon_sob}, we consider $F_\text{var}$. 
We note that the Sobol' indices for the $F_\text{var}$ take longer to converge
than for $F_\text{mean}$, for the present QoI.  However, even with a modest
number of MCMC samples (about 2500), the Sobol' indices provide the correct
ranking of importance. The total indices converge with around $10^4$ MCMC
samples.

The numerical studies for the present model linear inverse problem provide a
proof-of-concept study of~\cref{alg:spelm}.  In particular,  availability of
analytic expressions for the HS mappings enables testing the accuracy of the
computed results. We note that a modest MCMC sample size is sufficient to 
obtain the correct parameter rankings. We also observe that
fewer MCMC samples are required to estimate the indices $F_\text{mean}$ compared
to $F_\text{var}$.  This is not surprising, because computing second order
moments typically require more effort than that required for computing the mean.

\subsection{Nonlinear Bayesian inverse problem based on SEIR model}\label{sec:seir}
In this section, we consider a Bayesian inverse problem governed by the susceptible-exposed-infected-recovered (SEIR) model~\cite{SEIR00,ModernEp}
epidemic model.  In~\cref{sec:SEIR_model} we discuss the governing SEIR model
and the Bayesian inverse problem under study.  In~\cref{sec:SEIR_sampling},
we study the proposed importance sampling procedure for computing the HS
mappings under study.  Finally, in~\cref{sec:SEIR_GSA}, we present our
computational results for GSA of the present Bayesian inverse problem with
respect to prior hyperparameters.

\subsubsection{The inverse problem}\label{sec:SEIR_model}
The SEIR model simulates the time dynamics of an epidemic outbreak in a  
population. 
The model has four
compartments, $S$, $E$, $I$, and $R$, corresponding to the
susceptible, exposed, infected, and recovered populations.  The
individuals in the exposed compartment are those who have been exposed to the
disease but are not yet displaying signs of infection. The
individuals in the $I$ compartment are infected and infectious.  We consider a standard SEIR model where we assume recovered individuals
cannot be reinfected.  Additionally, we assume that the natural birth and death
rates are equal and neglect disease related mortality. This ensures that the
total population $N=S+E+I+R$ remains constant over time.
The present model is described by the following system of
nonlinear ordinary differential equations (ODEs):
\begin{equation}\label{equ:seir}
\begin{aligned}
\dot{S} &=  \mu N - \beta SI/N - \mu S, \\
\dot{E} &=  \beta SI/N - (\sigma + \mu)E, \\
\dot{I} &= \sigma E - (\gamma + \mu) I, \\
\dot{R} &=  \gamma I - \mu R.
\end{aligned}
\end{equation}
There are four model parameters in the above system which we seek to
estimate. The infection rate $\beta$, in units days$^{-1}$, represents how
quickly an infected individual infects a susceptible individual. The recovery
rate $\gamma$, in units days$^{-1}$, represents how fast an infected individual
recovers from infection. The latency rate $\sigma$, in units days$^{-1}$,
represents how long it takes for an exposed individual to display symptoms.
Lastly, there is also a parameter $\mu$, with units individuals per day, which
represents both the natural birth rate and the natural death rate. In the model,
individuals are only born susceptible while individuals in any compartment can die a natural death.
As noted before, since the birth and death rates are the same, the total
population size remains constant. 

\begin{figure}[h!!]
    \centering
   \subfigure{\includegraphics[width=.45\textwidth]{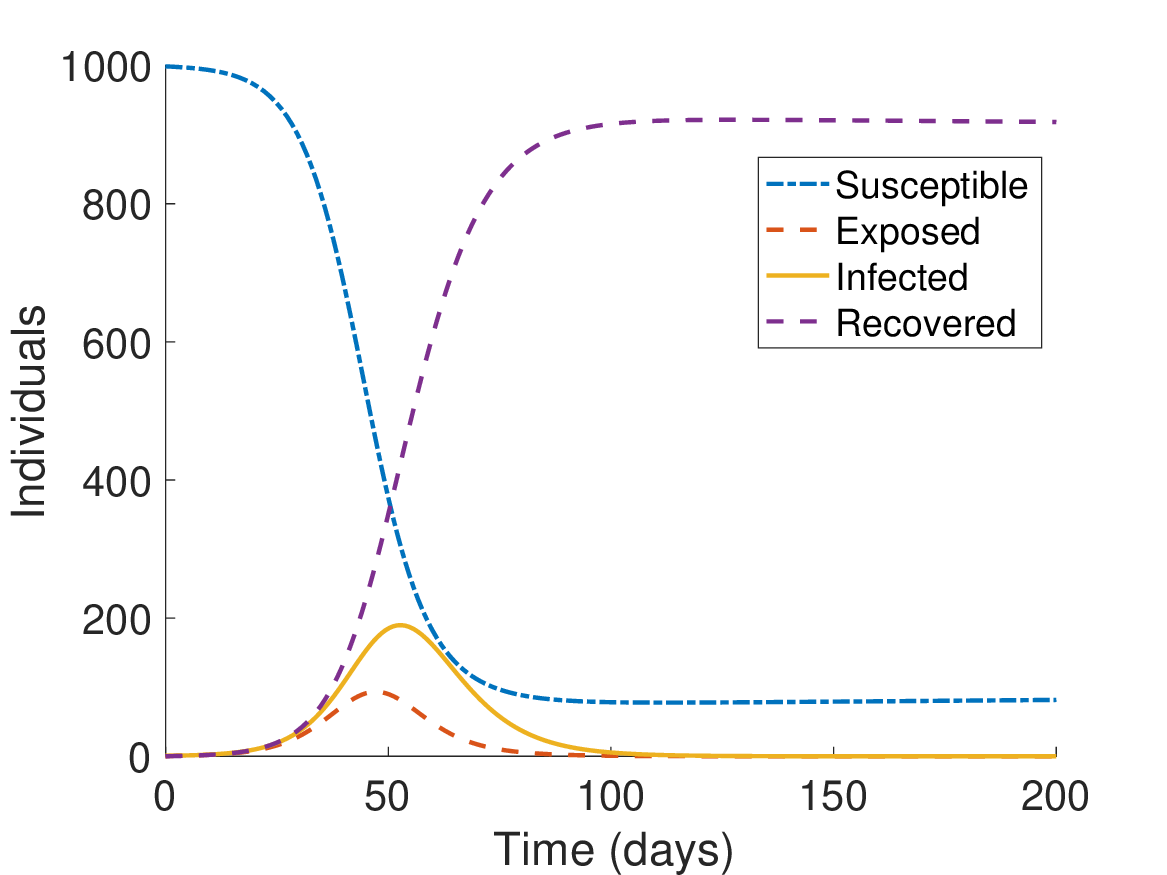}}
      \subfigure{\includegraphics[width=.45\textwidth]{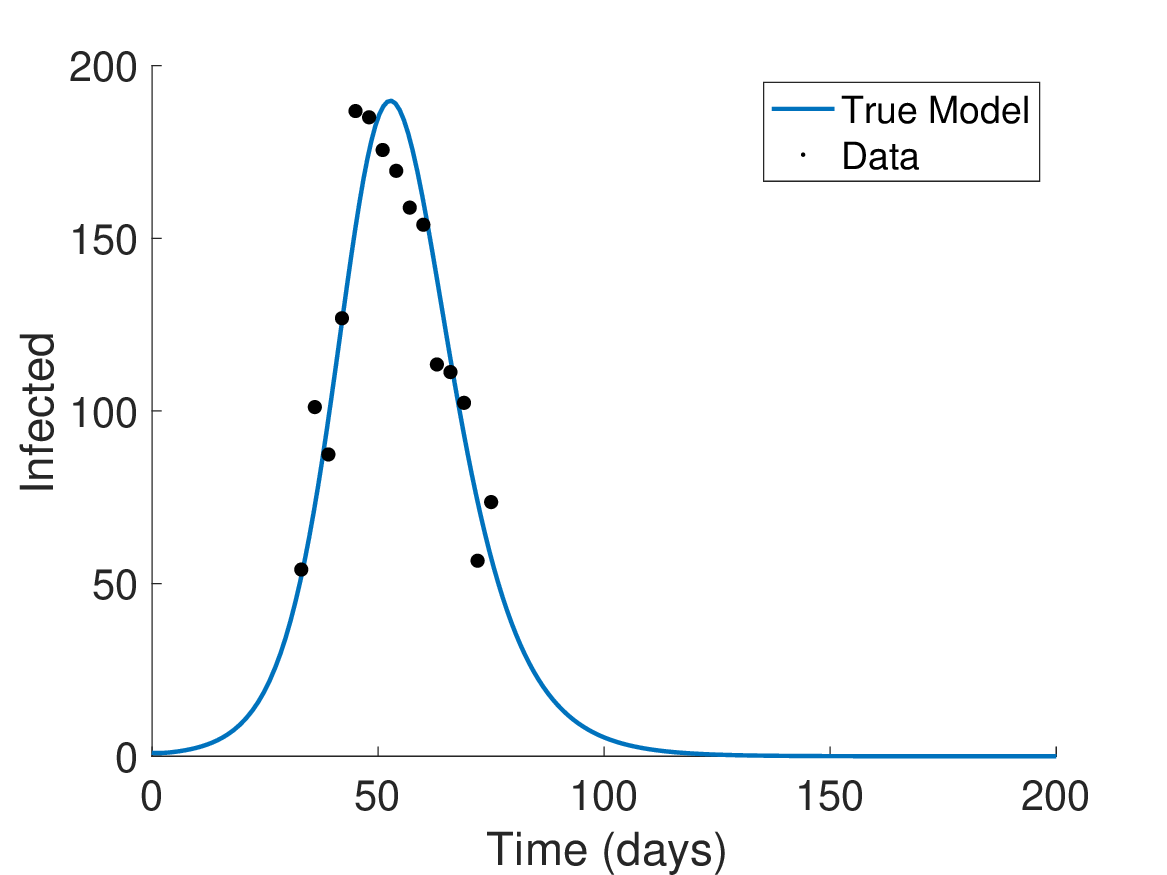}}
    \caption{Left: Simulated dynamics of an epidemic outbreak following the SEIR model. The total population is $N=1000$ individuals and the model parameters are $\mu=5.48\times 10^{-5}$ individuals$/$day, $\beta = 1/2.5$ days$^{-1}$, $\sigma=1/3$ days$^{-1}$, and $\gamma = 1/7$ days$^{-1}$. Initially, there is one infected individuals and there are no exposed individuals. Right: Simulated data of the epidemic outbreak compared to the true infected dynamics. The simulated infected data is taken by adding noise to values from the true model. The noise is sampled from the normal distribution $\mathcal{N}(0,30^2)$.}
    \label{fig:dyn}
\end{figure}

\textbf{Setup.}
For the purposes of this example, we simulate an epidemic governed by the SEIR
model for a population of $N=1000$ individuals. The nominal parameters and
initial conditions are detailed in~\cref{tab:seir}.  The nominal parameter
values will be used as ``ground-truth'' in the computational studies that
follow. 
\begin{table*}[h!]
\centering
\begin{tabular}{cc|cc}
 Model Parameter & Value  & Initial Condition & Value  \\
\hline
$\mu$ & $5.48\times 10^{-5} $  & $S_0$ & $999$  \\
$\beta $ &        $1/2.5$     &  $E(0)$  &   $0$     \\
$\sigma $ &    $1/3$          &  $I(0)$  &    $1$      \\
$\gamma $ &      $1/7$         &  $R(0)$  &     $0$   
\end{tabular}
\vspace{1mm}
\caption{Model parameters and initial conditions used to simulate the SEIR model~\eqref{equ:seir} in~\cref{fig:dyn}.}\label{tab:seir}
\end{table*}
The dynamics of the epidemic under these conditions are shown in~\cref{fig:dyn}~(left).

Next, we formulate a Bayesian inverse problem.
In what follows, we formulate the inverse problem as that 
of estimating the log of the uncertain model parameters. 
Hence, we consider the inversion parameter vector,
$\vec\theta = \big[\begin{array}{cccc} 
   \log\mu & \log\beta & \log\sigma &
\log\gamma \end{array}\big]^\top$.  
The data measurements, used to solve the inverse problem, 
consist of simulated data $\{(t_k,I_k)\}$ at times
$t_k=3k+30,$ where $k=1,\ldots,15$.  These simulated data measurements are
obtained by  solving the SEIR model with ground-truth parameter values and
adding random noise.
The
noise at each measurement is identically independently distributed from a normal
distribution $\mathcal{N}(0,30^2)$. The simulated data compared to the true
model are shown in~\cref{fig:dyn}~(right). 

We use a Gaussian prior 
$\mathcal{N}(\vec{m}_\text{pr}, \vec\Sigma_\text{pr})$
on the inversion parameter vector 
$\boldsymbol\theta$ with
\begin{equation}\label{equ:pr_seir}
\vec{m}_\text{pr} = \left[\begin{array}{c}
m_{\log\mu} \\
m_{\log\beta} \\
m_{\log\sigma} \\
m_{\log\gamma}
\end{array}\right], \quad 
\vec\Sigma_\text{pr} = \left[\begin{array}{cccc}
s_{\log\mu}^2 & 0 & 0 & 0 \\
0 & s_{\log\beta}^2  & 0 & 0 \\
0 & 0  & s_{\log\sigma}^2 & 0 \\
0 & 0  & 0 & s_{\log\gamma}^2 
\end{array}\right].
\end{equation}
Note that, unlike the inverse problem in~\cref{sec:linear}, this Bayesian
inverse problem is nonlinear. In this case, we do not have access to an
analytically known posterior distribution.  This means Markov Chain Monte Carlo
(MCMC) is needed to sample from the posterior distribution. 

\textbf{Uncertainty in prior hyperparameters.}
We assume there is uncertainty in the hyperparameters that appear in~\eqref{equ:pr_seir}. 
Specifically, we consider the vector 
\[
\vec\xi = \big[\begin{array}{cccccccc}
m_{\log\mu}  & m_{\log\beta}  & m_{\log\sigma} & m_{\log\gamma} &  s_{\log\mu}^2  
   & s_{\log\beta}^2 & s_{\log\sigma}^2 & s_{\log\gamma}^2 \end{array}\big]^\top\]
of parameters that define the prior as uncertain.  In the present study, we
assume that the entries of $\vec\xi$ are independent uniformly distributed
random variables, as specified in \cref{tab:seir_unc}.
\begin{table*}[h!]
\centering
\begin{tabular}{c|c|c|c}
Mean Hyperparameter &  Distribution & Variance Hyperparameter & Distribution \\
\hline
$m_{\log\mu} $ & $\mathcal{U}([-15,-5])$ & $s_{\log\mu}^2$ & $\mathcal{U}([0.5,1.5])$ \\
$m_{\log\beta} $ & $\mathcal{U}([-2.25,-0.75])$  & $s_{\log\beta}^2 $ & $\mathcal{U}([0.5,1.5])$  \\
$m_{\log\sigma} $ & $\mathcal{U}([-2.25,-0.75])$   & $s_{\log\sigma}^2$ & $\mathcal{U}([0.5,1.5])$   \\
$m_{\log\gamma}$ & $\mathcal{U}([-2.25,-0.75])$  & $s_{\log\gamma}^2$ & $\mathcal{U}([0.5,1.5])$  
\end{tabular}
\vspace{1mm}
\caption{Intervals for admissible hyperparameter values of the prior 
$\boldsymbol\theta\sim\mathcal{N}(\vec{m}_\text{pr},\vec\Sigma_\text{pr})$. 
Each hyperparameter is uniformly distributed on an interval perturbed $\pm 50\%$ 
of the respective nominal value. }\label{tab:seir_unc}
\end{table*} 

\textbf{Quantity of interest.}
An important quantity of interest in epidemiology is the basic reproduction number, denoted $R_0$. 
It can be interpreted as the number of secondary infections caused, on average, by a single 
individual~\cite{SEIR00}. Determining $R_0$ of an epidemic is key to understanding 
how severe the outbreak could be. For the SEIR model~\cref{equ:seir},  $R_0$ takes the form
\begin{equation}\label{equ:r0}
R_0=\frac{\beta}{\gamma + \mu}\frac{\sigma}{\sigma+\mu}.
\end{equation} 
For the epidemic in~\cref{fig:dyn}, $R_0=2.7985$.  The importance of $R_0$ makes
it a prime area to apply uncertainty quantification and robustness analysis.
In~\cite{Lloyd09}, the robustness of $R_0$ estimates to model parameters is
considered through local derivative-based methods. Hence, we focus on $R_0$
as the QoI,
\[
q(\vec\theta) = 
\frac{e^{\theta_2}}{e^{\theta_4} + e^{\theta_1}}
\frac{e^{\theta_3}}{e^{\theta_3}+e^{\theta_1}}.
\]

\subsubsection{Parameter estimation and importance sampling}
\label{sec:SEIR_sampling}
Before we can implement~\cref{alg:spelm}, we have to choose the importance
sampling distribution. In accordance with the discussion in~\cref{sec:method},
we choose the importance sampling distribution $\pi_\text{pr}^\text{IS}$ as
$\mathcal{N}(\vec{m}_\text{pr}^\text{IS},\vec{\Sigma}_\text{pr}^\text{IS})$
with 
\begin{equation}\label{equ:is_pr}
\vec{m}_\text{pr}^\text{IS}= \left[\begin{array}{cc}
-10 \\
-1.5\\
-1.5 \\
-1.5
\end{array}\right], \quad \boldsymbol\Sigma_\text{pr}^\text{IS} = \left[\begin{array}{cccc}
3^2 & 0 & 0 & 0 \\
0 & 2^2 & 0 & 0\\
0 & 0 & 2^2 & 0 \\
0 & 0 & 0 & 2^2
\end{array}\right].
\end{equation}

\begin{figure}[h!]
    \centering
       \subfigure{\includegraphics[width=0.24\textwidth]{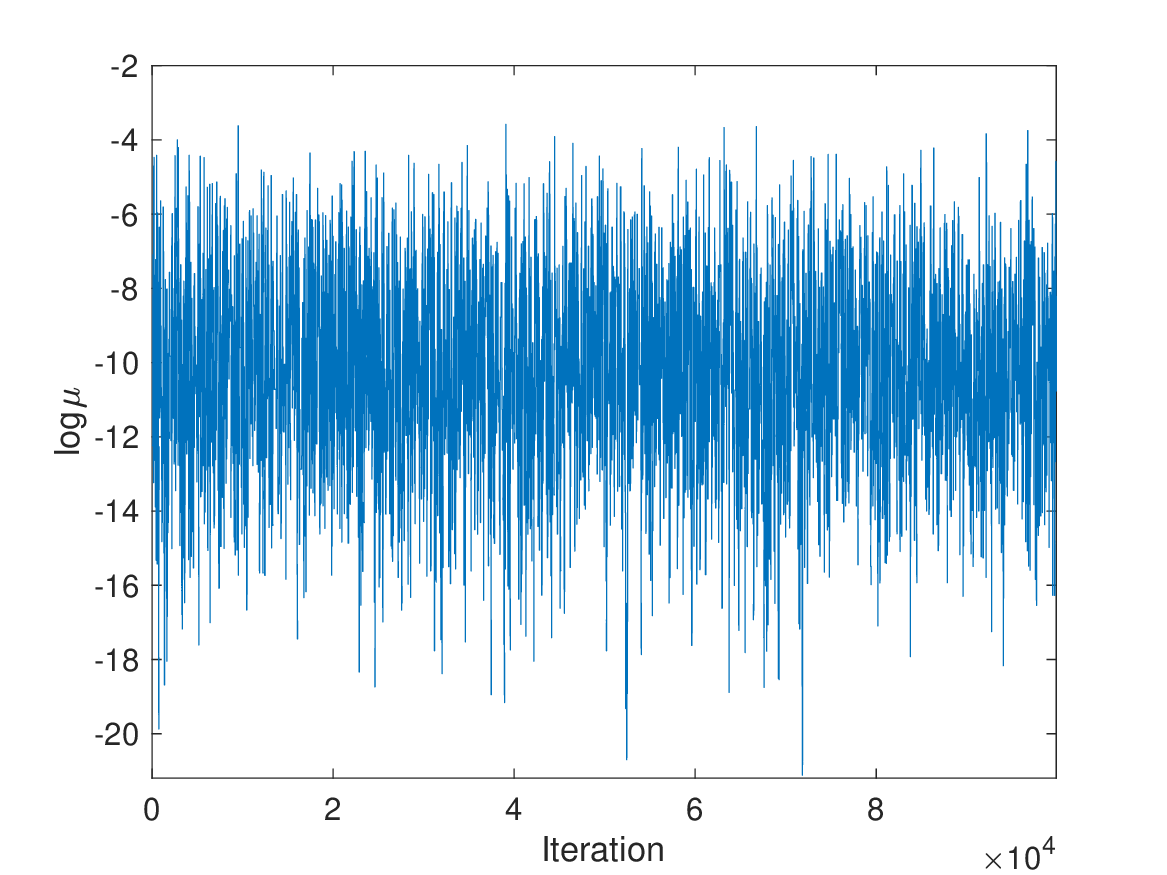}}
   \subfigure{\includegraphics[width=0.24\textwidth]{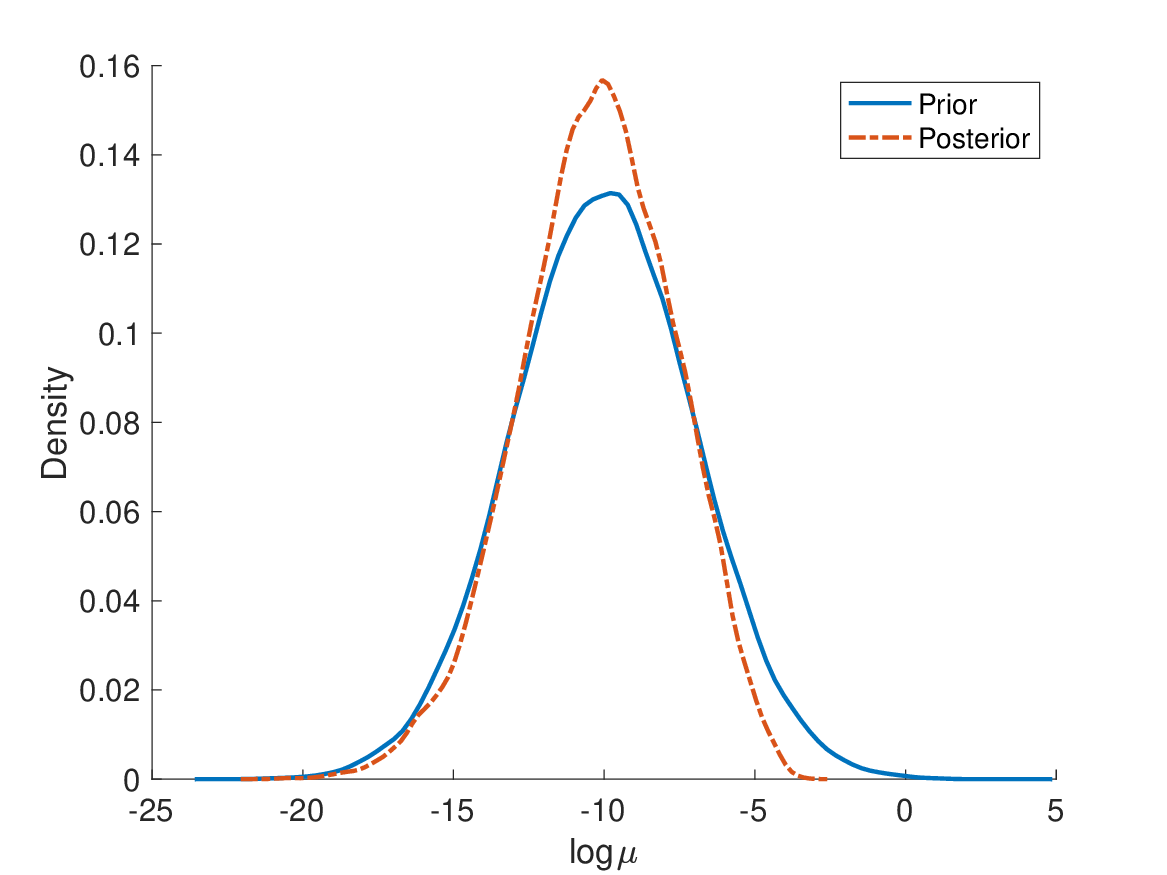}}
   \subfigure{\includegraphics[width=0.24\textwidth]{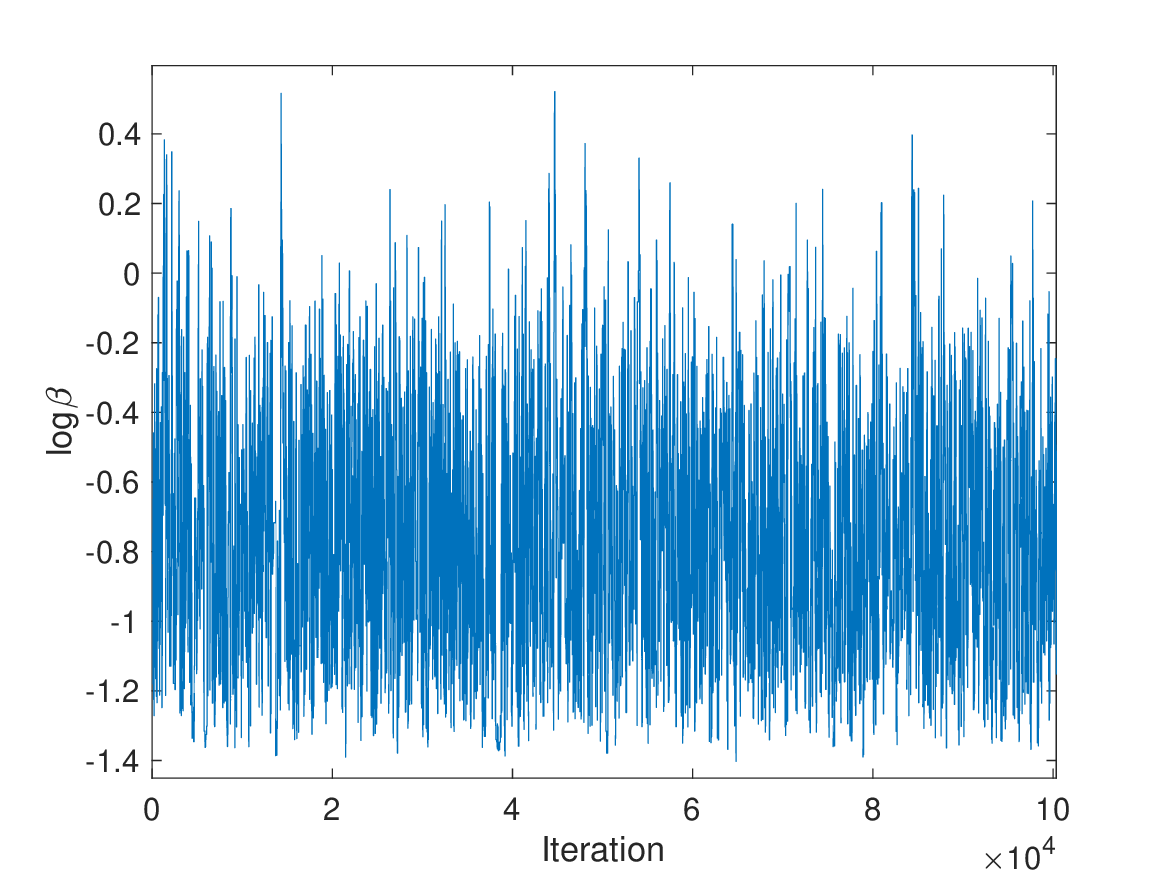}}
   \subfigure{\includegraphics[width=0.24\textwidth]{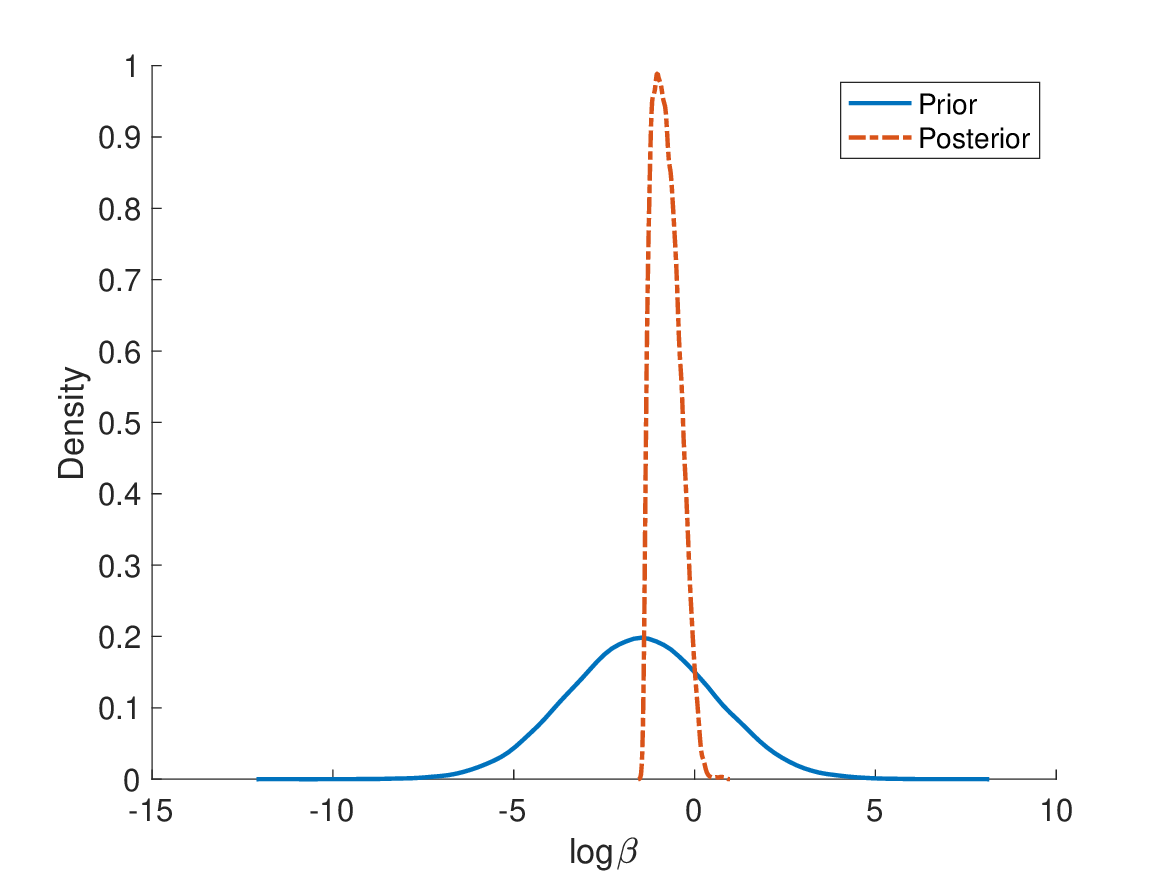}}
        \subfigure{\includegraphics[width=0.24\textwidth]{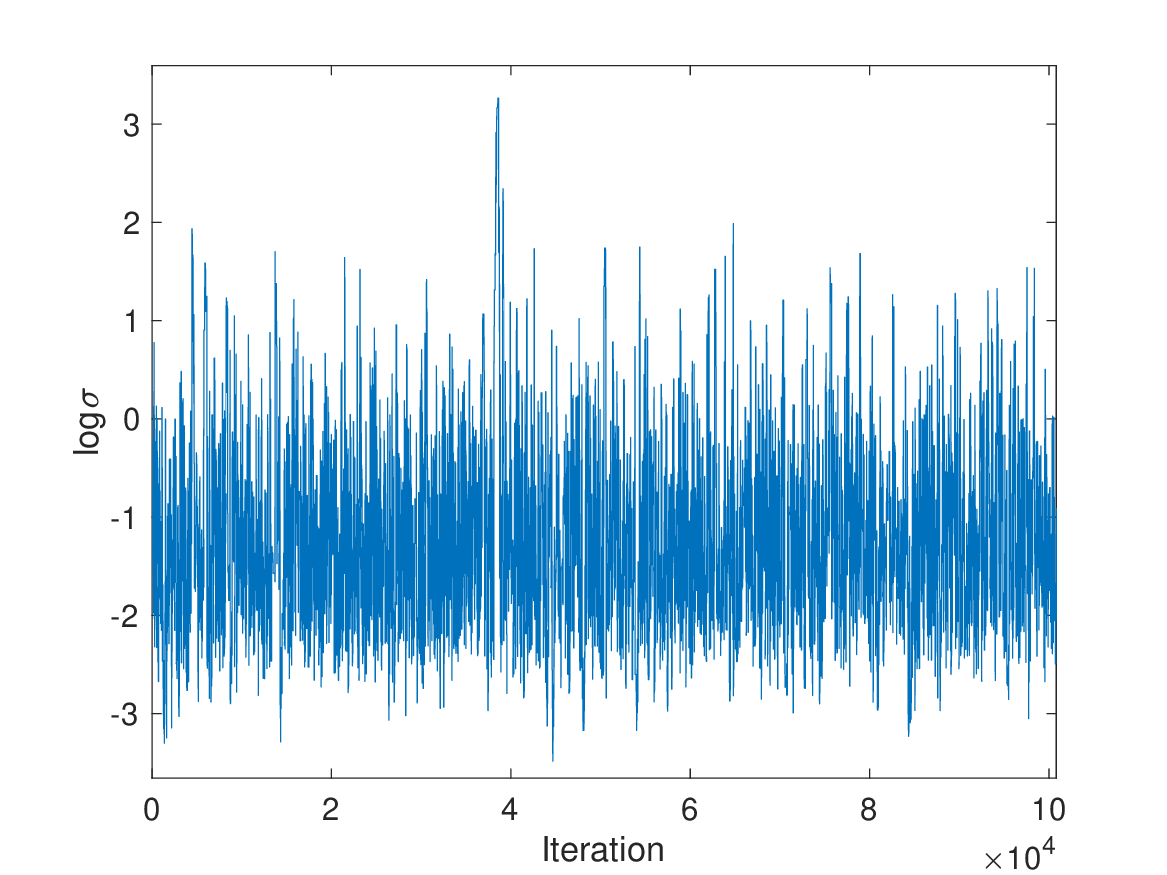}}
   \subfigure{\includegraphics[width=0.24\textwidth]{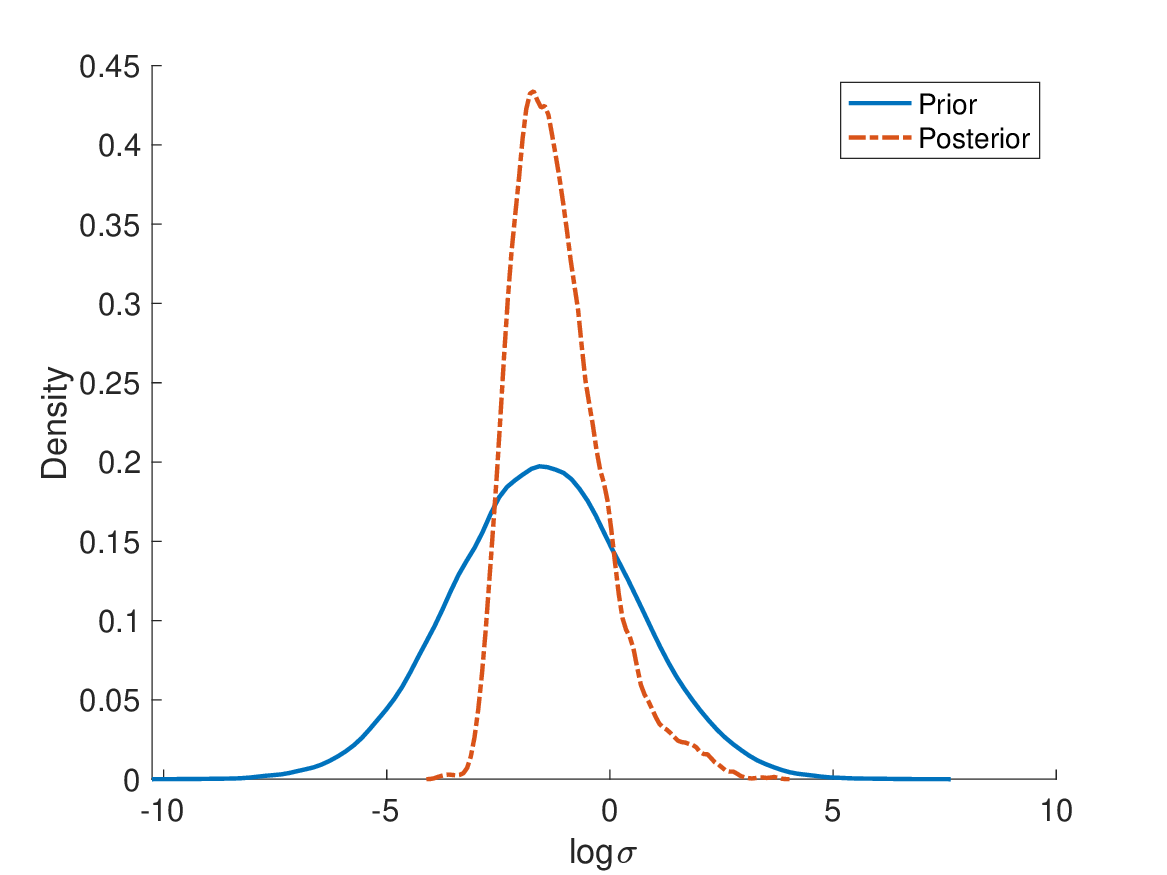}}
     \subfigure{\includegraphics[width=0.24\textwidth]{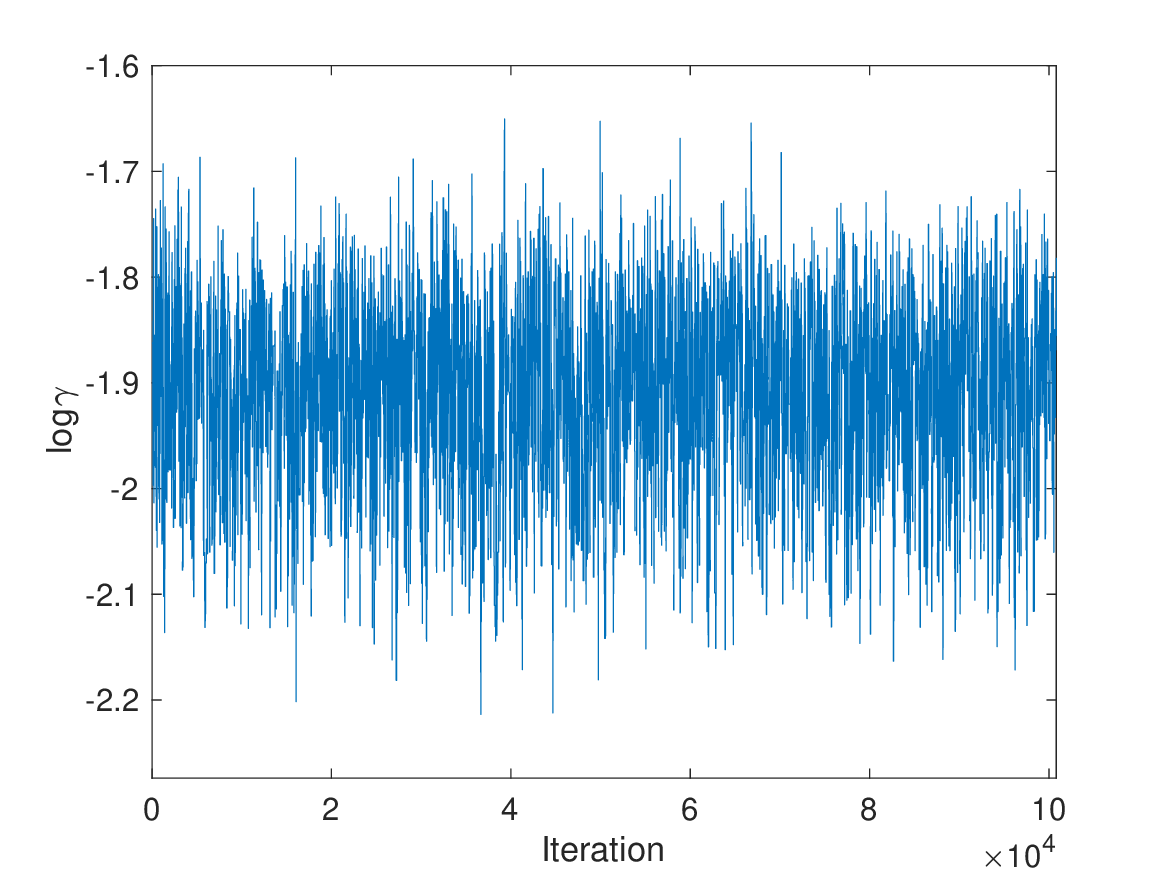}}
   \subfigure{\includegraphics[width=0.24\textwidth]{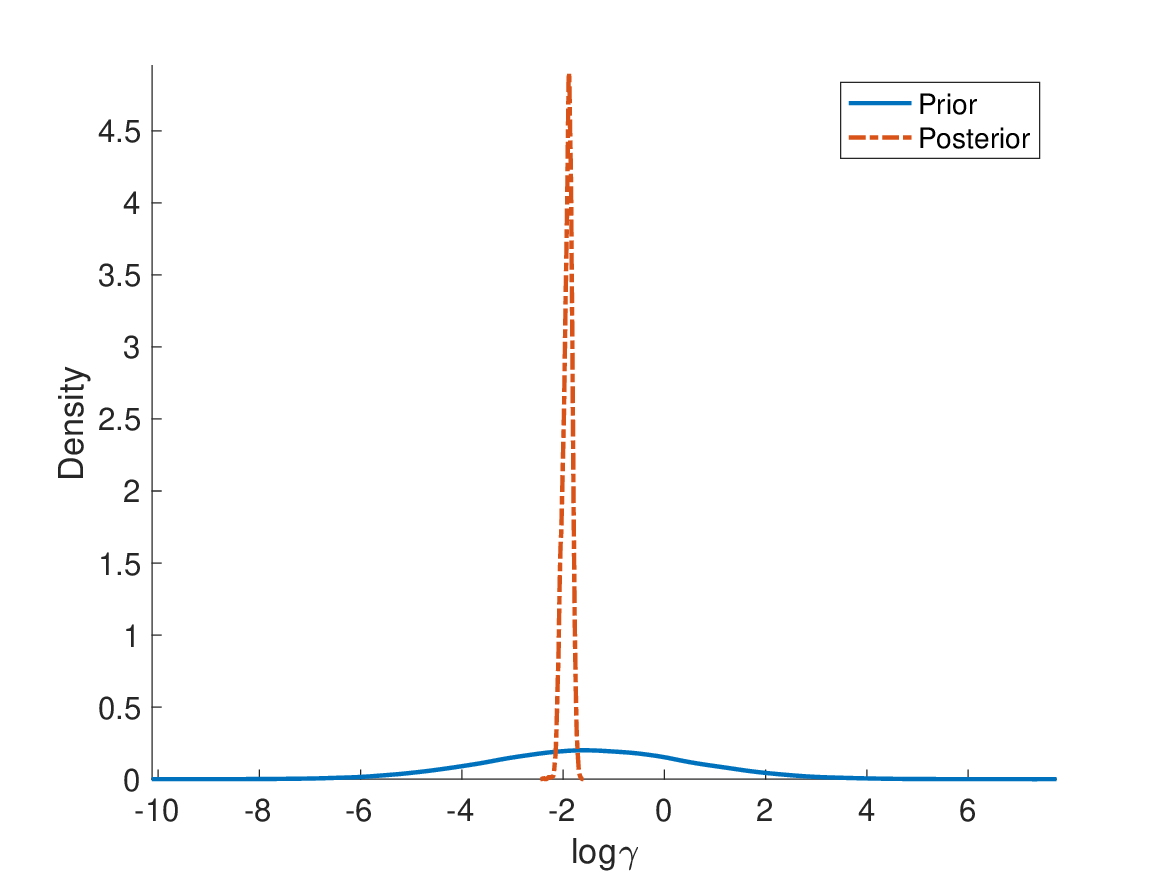}}
    \caption{Results of Bayesian parameter estimation for SEIR example.
     Chains are constructed by taking $1.5\times 10^5$ samples after $10^3$ iterations of burn-in.
      Marginal distributions are constructed by kernel density estimation on the respective MCMC chains.}
    \label{fig:seir_chains}
\end{figure}

Because $m_{\log\mu}$ takes a wider range of values compared to the other means, 
we impose a large variance on $\log\mu$ in $\pipris$. We construct the corresponding 
posterior $\pipostis$ using the DRAM algorithm. 
The first $10^3$ samples are removed for burn-in. After sufficient burn-in, we generate $1.5\times 10^5$ from the posterior. We present the MCMC chains of log parameters and their respective marginal posterior distributions in~\cref{fig:seir_chains}. 

\begin{figure}[h!]
    \centering
       \subfigure{\includegraphics[width=0.4\textwidth]{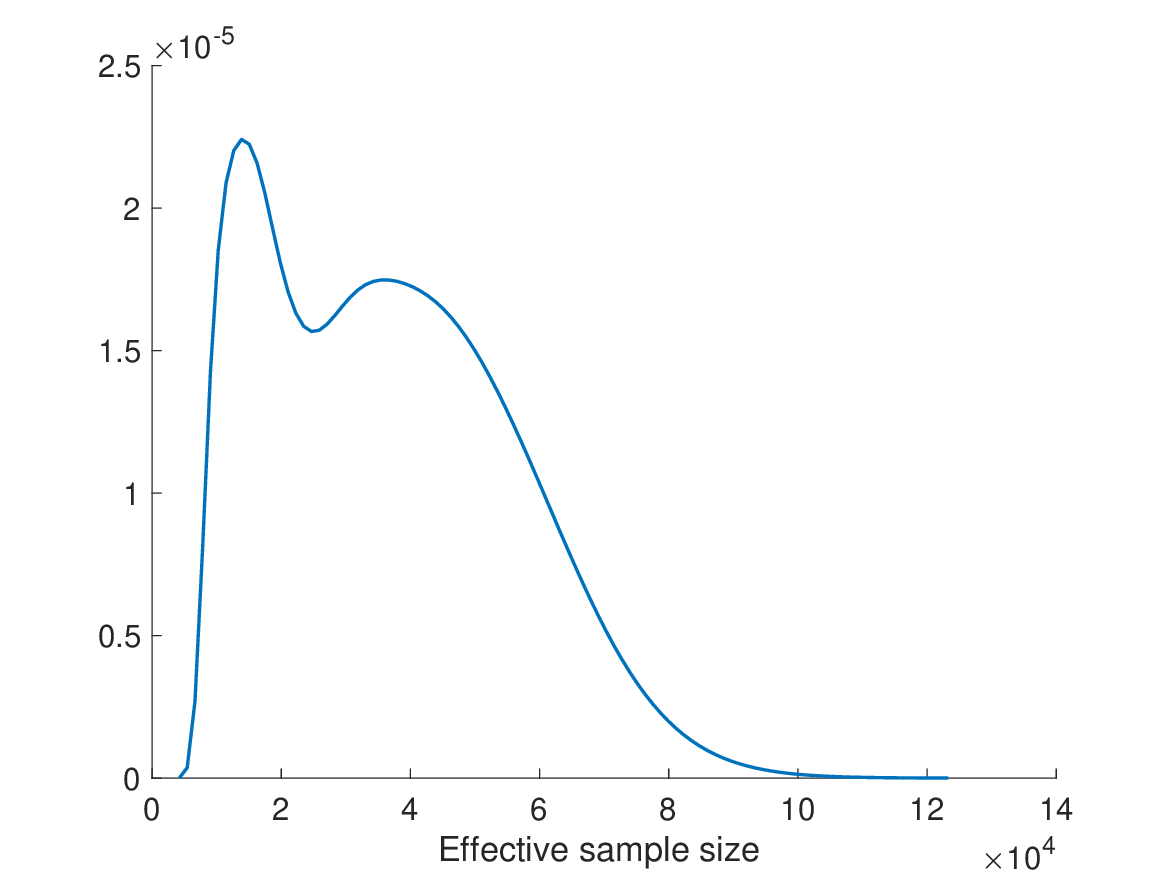}}
       \subfigure{\includegraphics[width=0.4\textwidth]{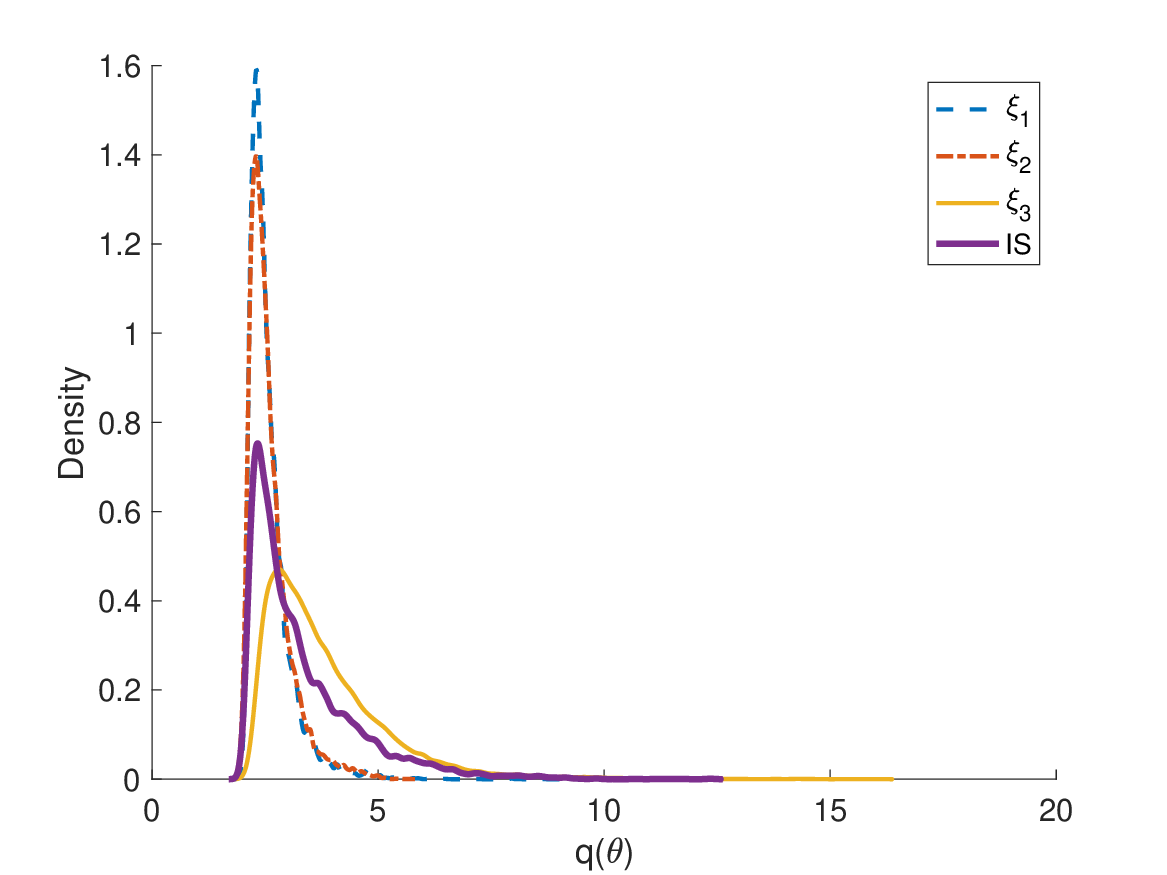}}  
    \caption{Left: Distribution of effective sample sizes using $\pipostis$. Right: Kernel density estimates of the distribution for $\pi_\text{post}^\text{IS}$ compared to those for selected target posterior distributions.}
    \label{fig:seir_diag}
\end{figure}

In~\cref{fig:seir_diag}~(left), we evaluate the effectiveness of our importance sampling distribution by 
examining the distribution of effective sample sizes.
We also compare the distribution of $R_0$ values, with respect to $\pi_\text{post}^\text{IS}$,
compared to the posterior distributions for three realizations of the prior hyperparameters
in~\cref{fig:seir_diag}~(right).
These results indicate that we can use $\pi_\text{post}^\text{IS}$ as an importance sampling
distribution for the target posteriors.

\subsubsection{Sensitivity analysis}\label{sec:SEIR_GSA}
Here, we study the sensitivity of the HS mappings $F_\text{mean}$,
$F_\text{var}$, and $F_\text{MAP}$ to prior
hyperparameters, relative  to the QoI $q(\vec\theta) = R_0$.  
As discussed in~\cref{sec:method}, $F_\text{MAP}$ is not evaluated the same way
as the other two HS mappings---it is evaluated by solving an optimization
problem. Therefore, we only include convergence studies for $F_\text{mean}$
and $F_\text{var}$.  

For each HS mapping, surrogate models are constructed using $10^3$ realizations
of $\vec\xi$, drawn using Latin hypercube sampling (LHS).  For polynomial chaos 
expansion surrogates, we use expansions of total degree 6.  SW-ELM surrogates
use $800$ realizations for training and $200$ for validation during the weight
sparsification step. 

\begin{figure}[h!]
    \centering  
       \subfigure[SW-ELM]{\includegraphics[width=0.4\textwidth]{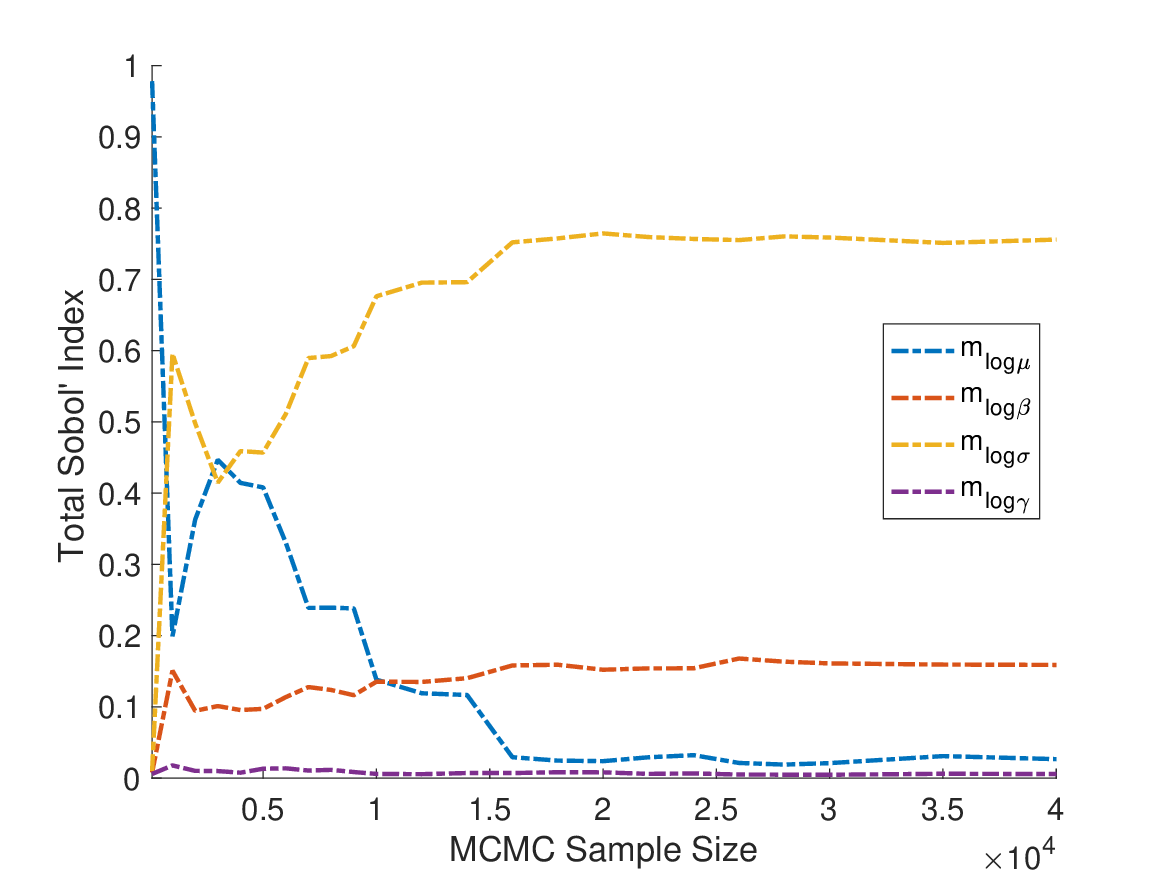}}
                \subfigure[SW-ELM]{\includegraphics[width=0.4\textwidth]{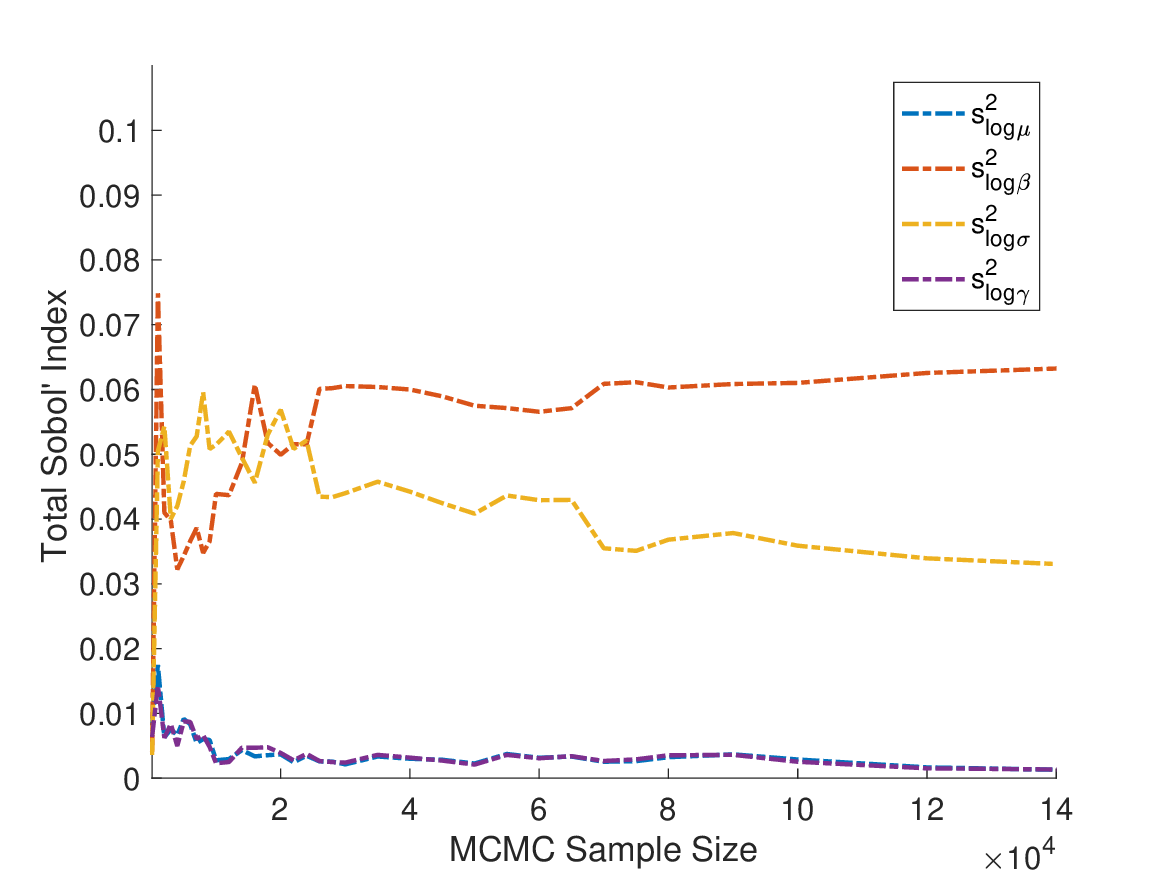}}
  \subfigure[PCE]{\includegraphics[width=0.4\textwidth]{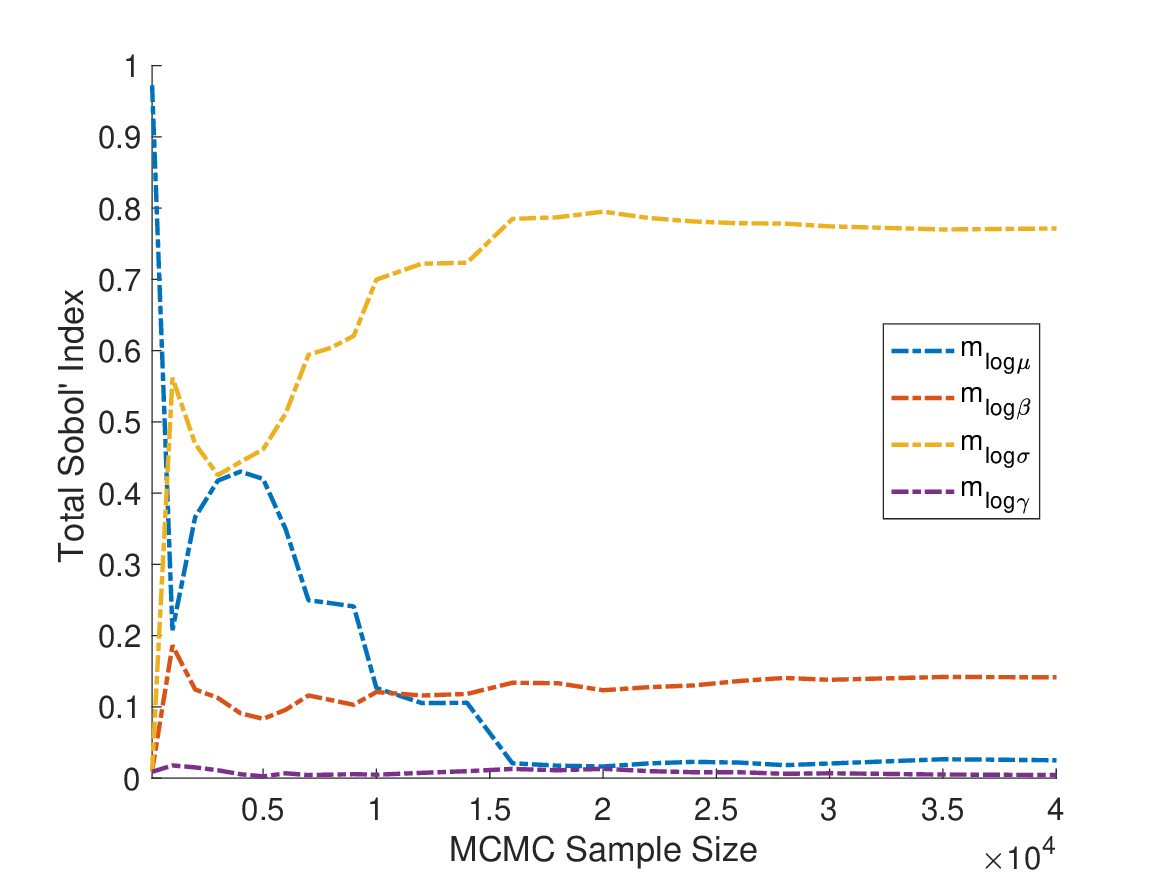}}
  \subfigure[PCE]{\includegraphics[width=0.4\textwidth]{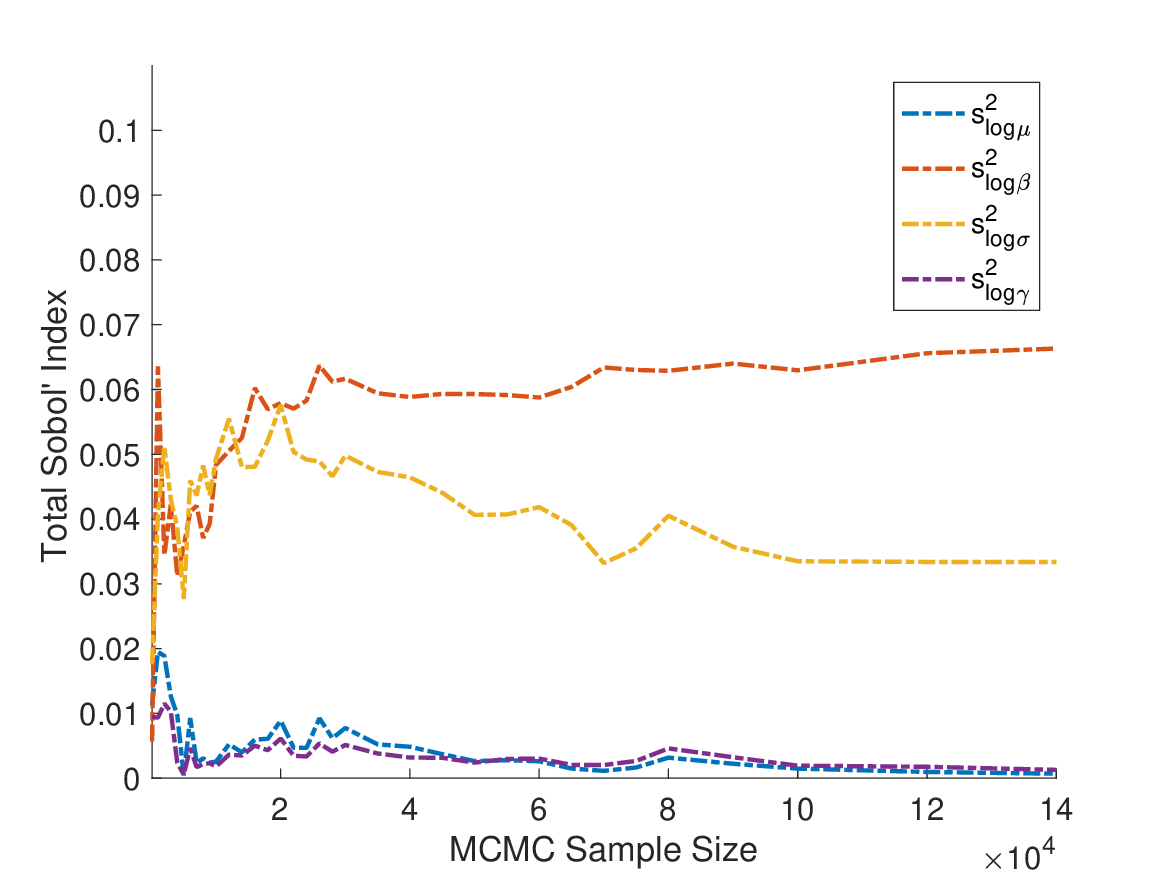}}
    \caption{Convergence of the total indices of $F_\mathrm{mean}(\boldsymbol\xi)$ with increasing MCMC sample size, comparing SW-ELM and PCE results. Total indices of the mean hyperparameters and variance hyperparameters are displayed separately.}
    \label{fig:tot_mu}
\end{figure}

\begin{figure}[h!]
    \centering  
       \subfigure[SW-ELM]{\includegraphics[width=0.4\textwidth]{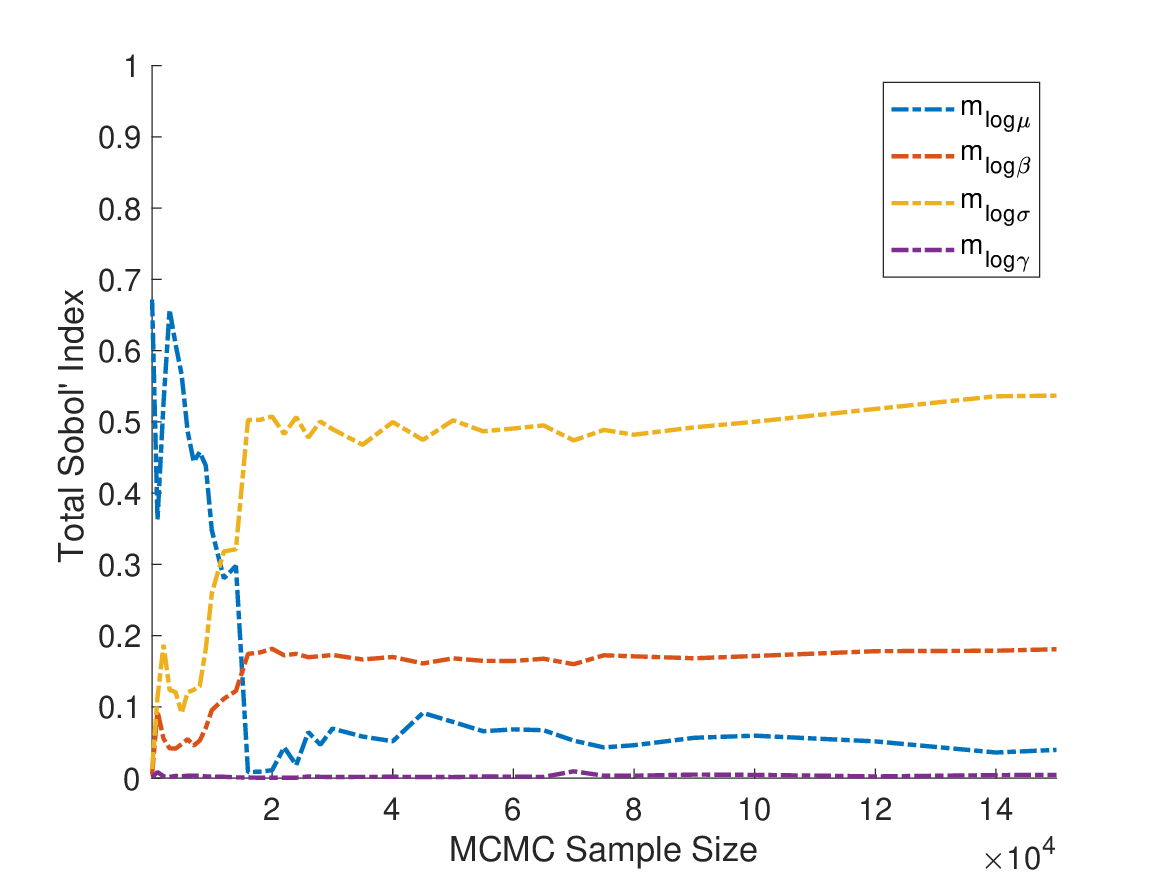}}
                \subfigure[SW-ELM]{\includegraphics[width=0.4\textwidth]{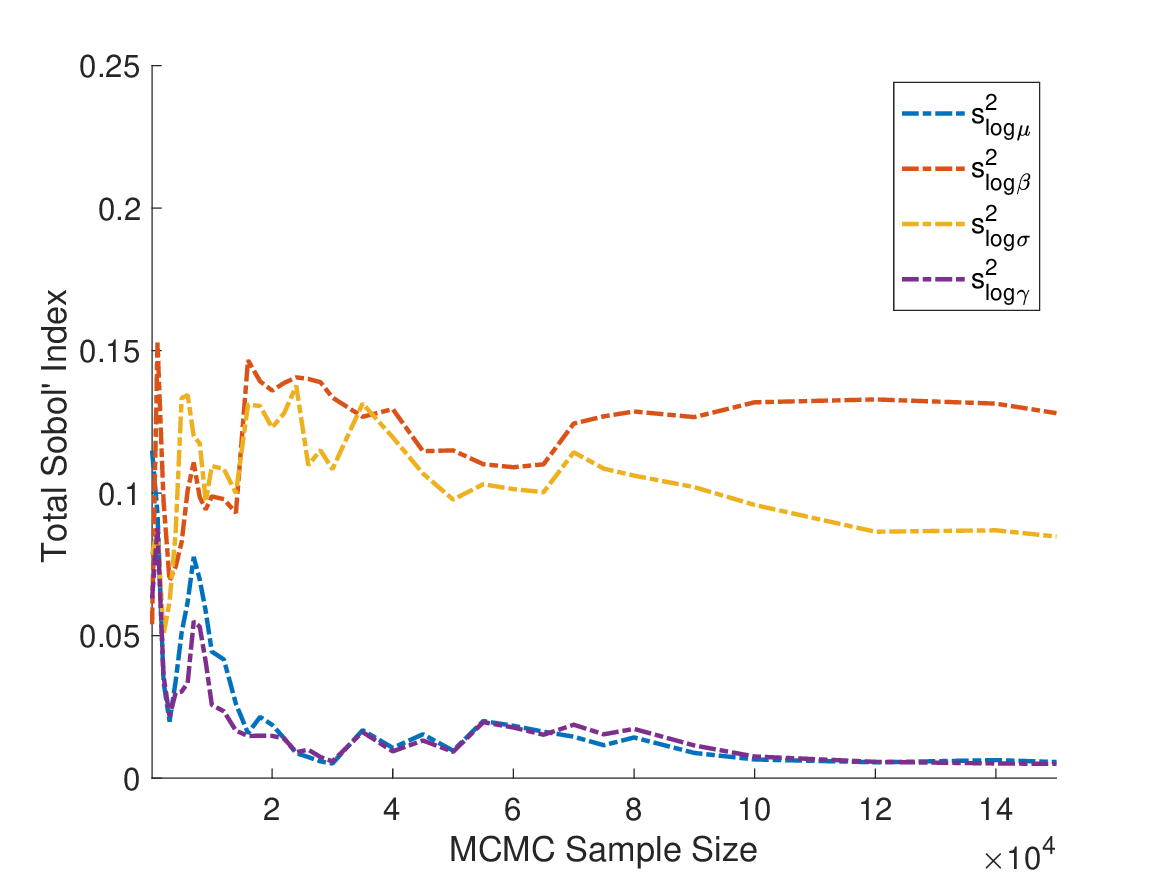}}
  \subfigure[PCE]{\includegraphics[width=0.4\textwidth]{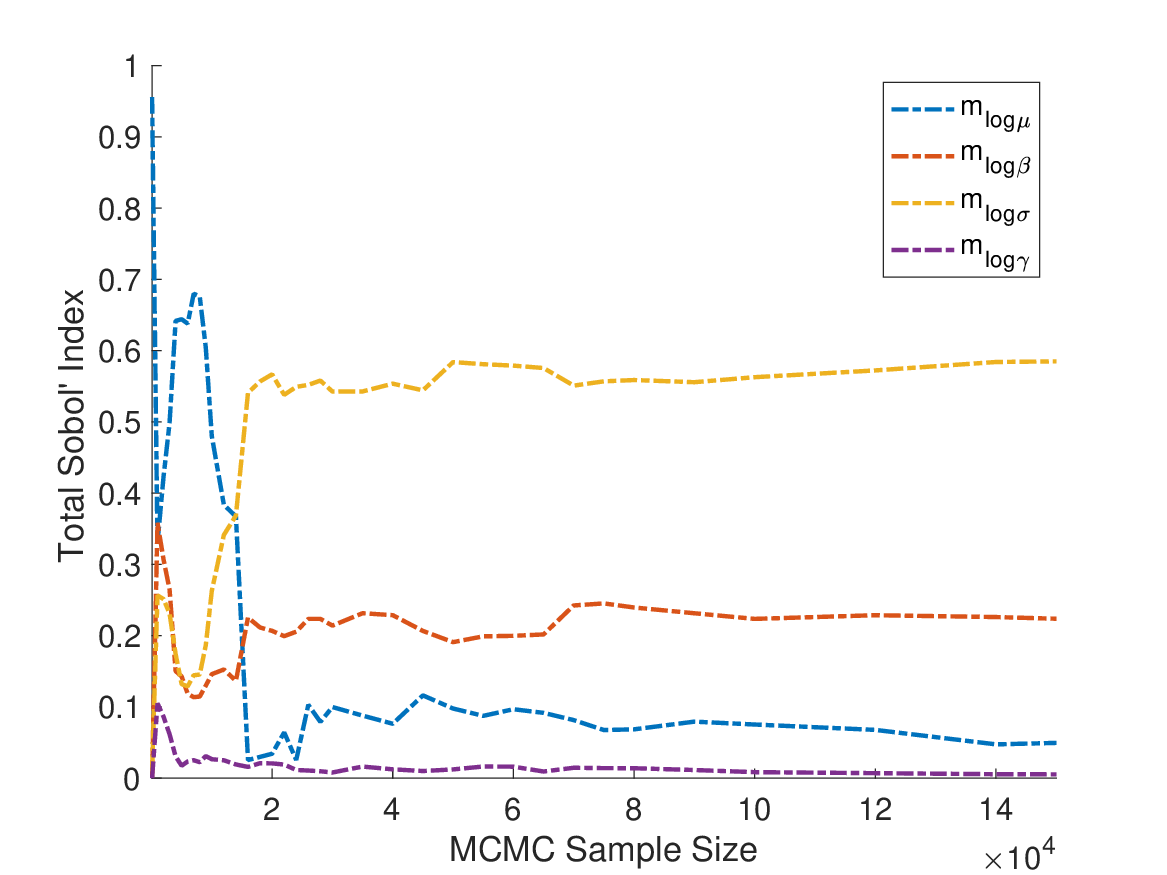}}
  \subfigure[PCE]{\includegraphics[width=0.4\textwidth]{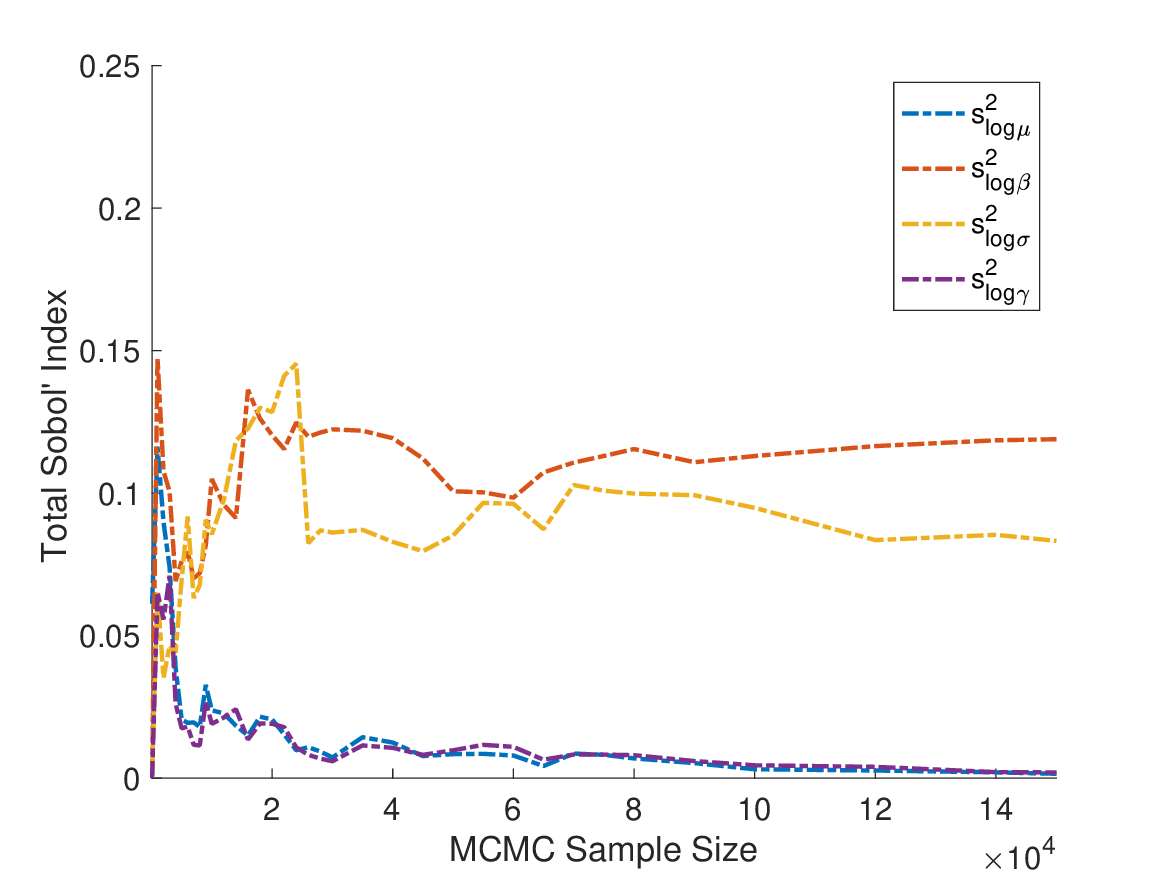}} 
    \caption{Convergence of the total indices of $F_\mathrm{var}(\boldsymbol\xi)$ with increasing MCMC sample size, comparing SW-ELM and PCE results. Total indices of the mean hyperparameters and variance hyperparameters are displayed separately.}
    \label{fig:tot_va}
\end{figure}

\begin{figure}[h!]
    \centering  
  \subfigure[Total Sobol' indices of $F_\text{mean}$]{\includegraphics[width=0.32\textwidth]{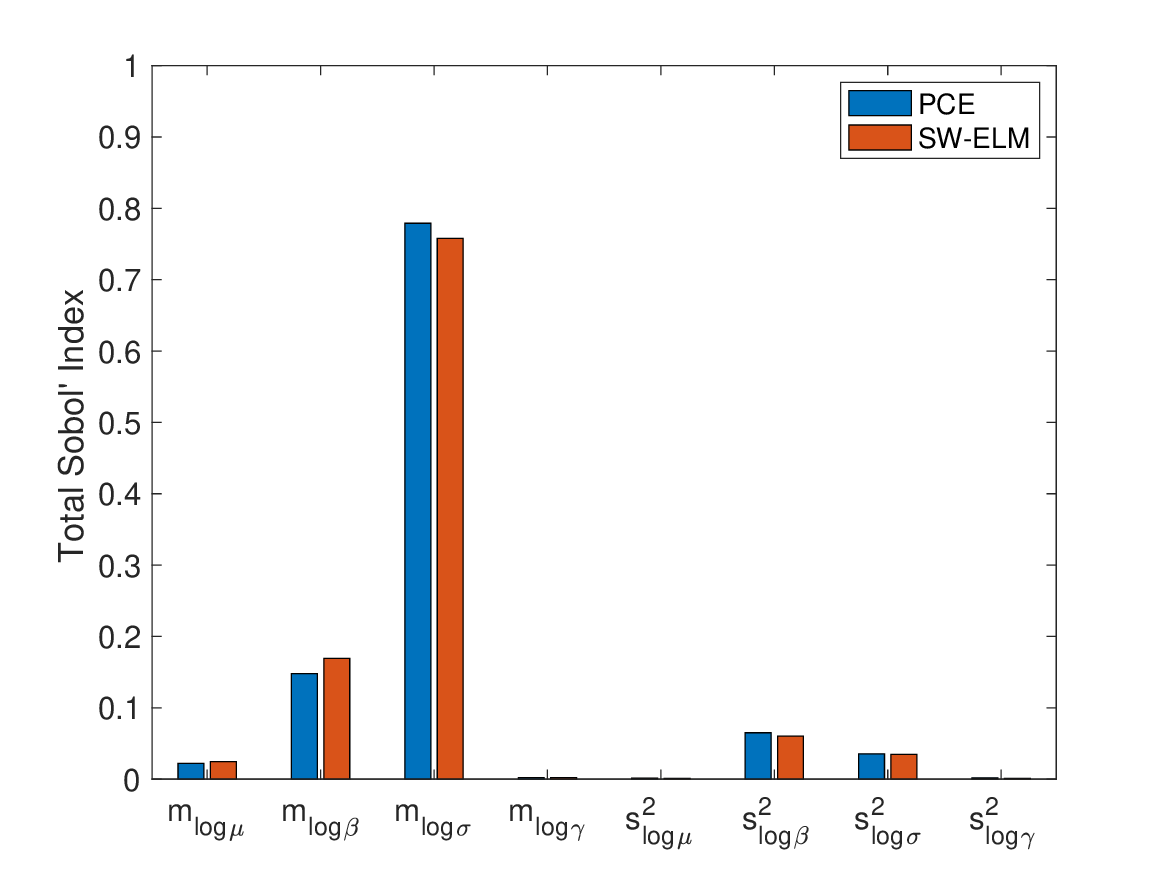}}  
  \subfigure[Total Sobol' indices of $F_\text{var}$]{\includegraphics[width=0.32\textwidth]{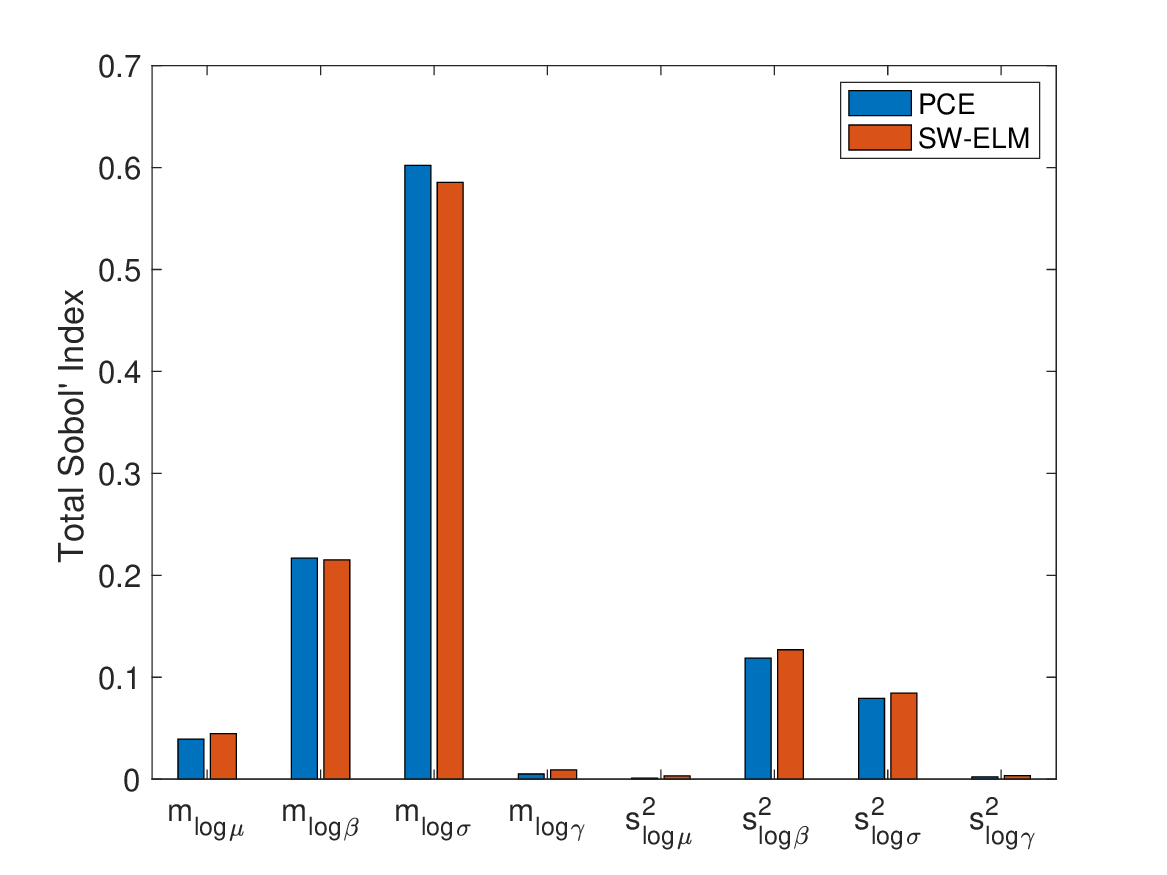}}  
  \subfigure[Total Sobol' indices of $F_\text{MAP}$]{\includegraphics[width=0.32\textwidth]{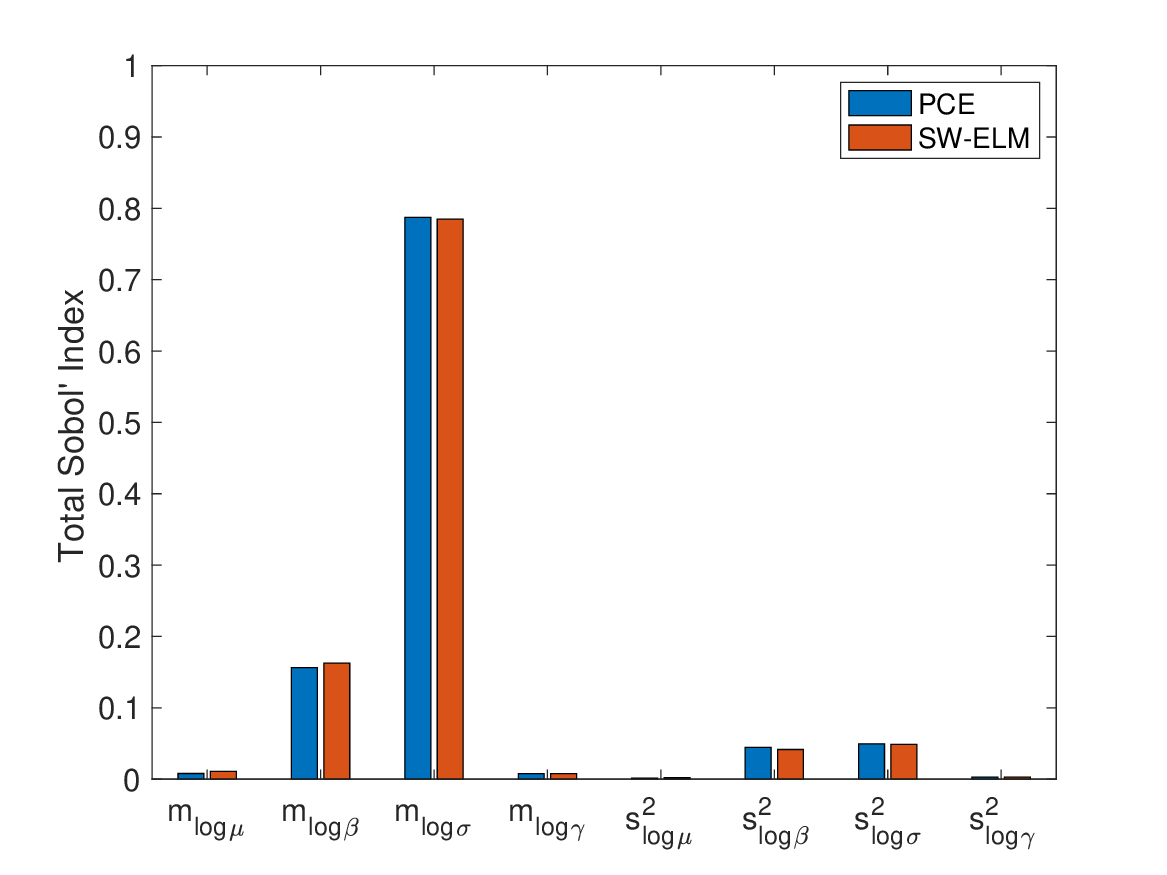}}
    \caption{Comparison of total Sobol' indices of $F_\text{mean}$, $F_\text{var}$, and $F_\text{MAP}$ estimated by SW-ELM and PCE surrogate-assisted GSA. $F_\text{mean}$ and $F_\text{var}$ are evaluated using $1.5\times 10^5$ MCMC samples drawn from $\pipostis$.}
    \label{fig:seir_bar}
\end{figure}

We start by studying the total Sobol' indices for 
$F_\mathrm{var}$ and $F_\mathrm{mean}$. We track the convergence of these
indices as 
we increase the
number of MCMC samples taken from $\pi_\text{post}^\text{IS}$ to up to $1.5\times
10^5$ samples.  The results are reported
in~\cref{fig:tot_mu}.   
We observe that the estimators for the larger Sobol' indices converge faster.
However, our importance ranking remains constant after $4\times 10^4$ MCMC
samples.  The convergence of the total indices of $F_\mathrm{var}$ are studied 
in~\cref{fig:tot_va}.   
As was observed when studying the linear Bayesian inverse problem, evaluating
the variance accurately requires more MCMC samples compared to evaluating the
mean.  Finally, we compare the converged total Sobol' indices of
$F_\mathrm{mean}(\boldsymbol\xi),F_\mathrm{var}(\boldsymbol\xi)$ with those of
$F_\mathrm{MAP}(\boldsymbol\xi)$ in~\cref{fig:seir_bar}. 

\begin{figure}[h!]
    \centering
       \subfigure[Mean]{\includegraphics[width=0.32\textwidth]{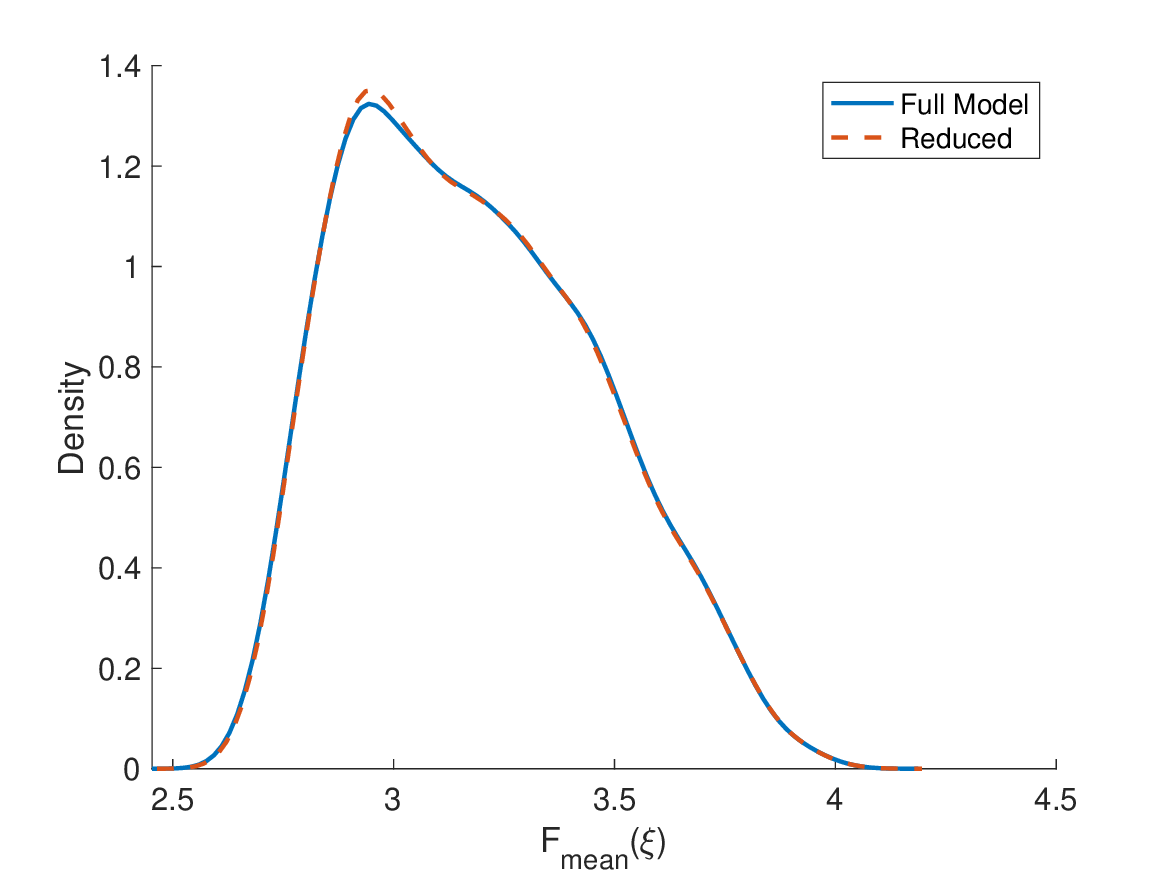}} 
    \subfigure[Variance]{\includegraphics[width=0.32\textwidth]{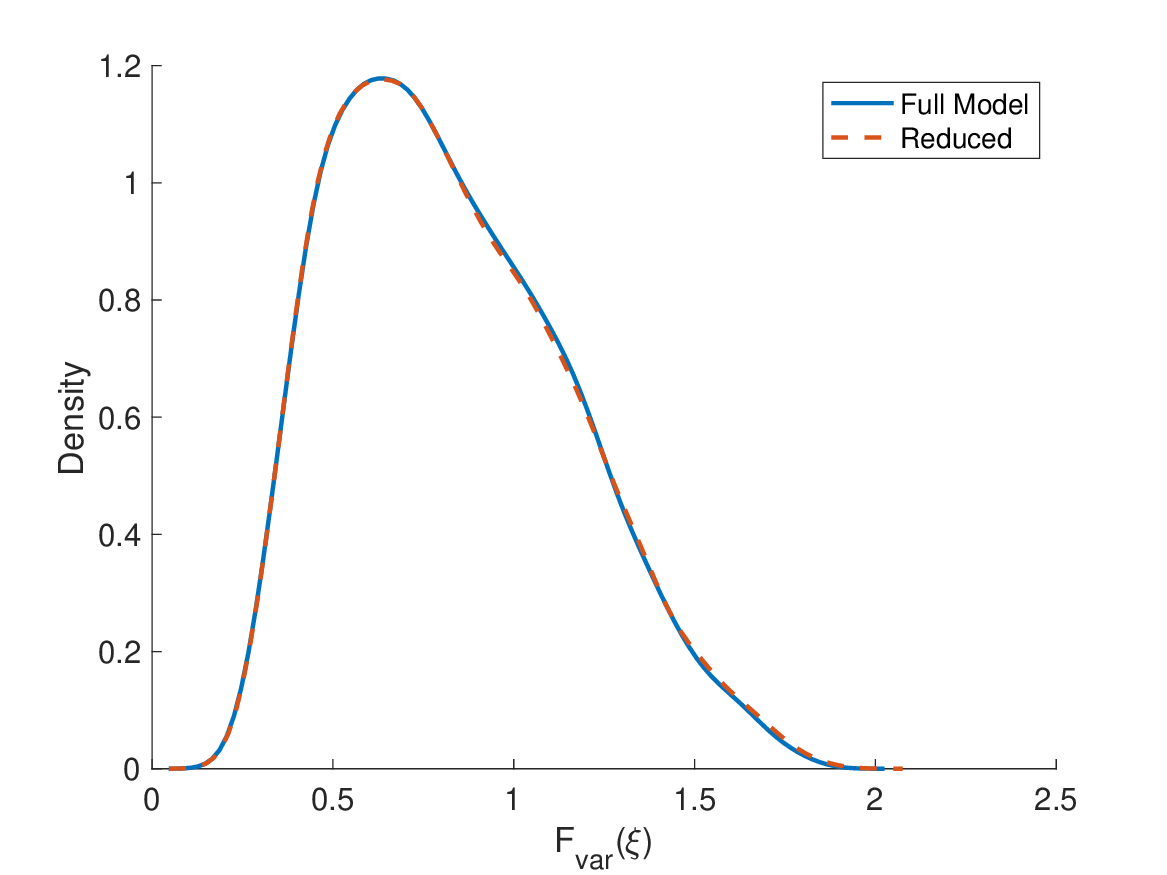}}
       \subfigure[MAP Point]{\includegraphics[width=0.32\textwidth]{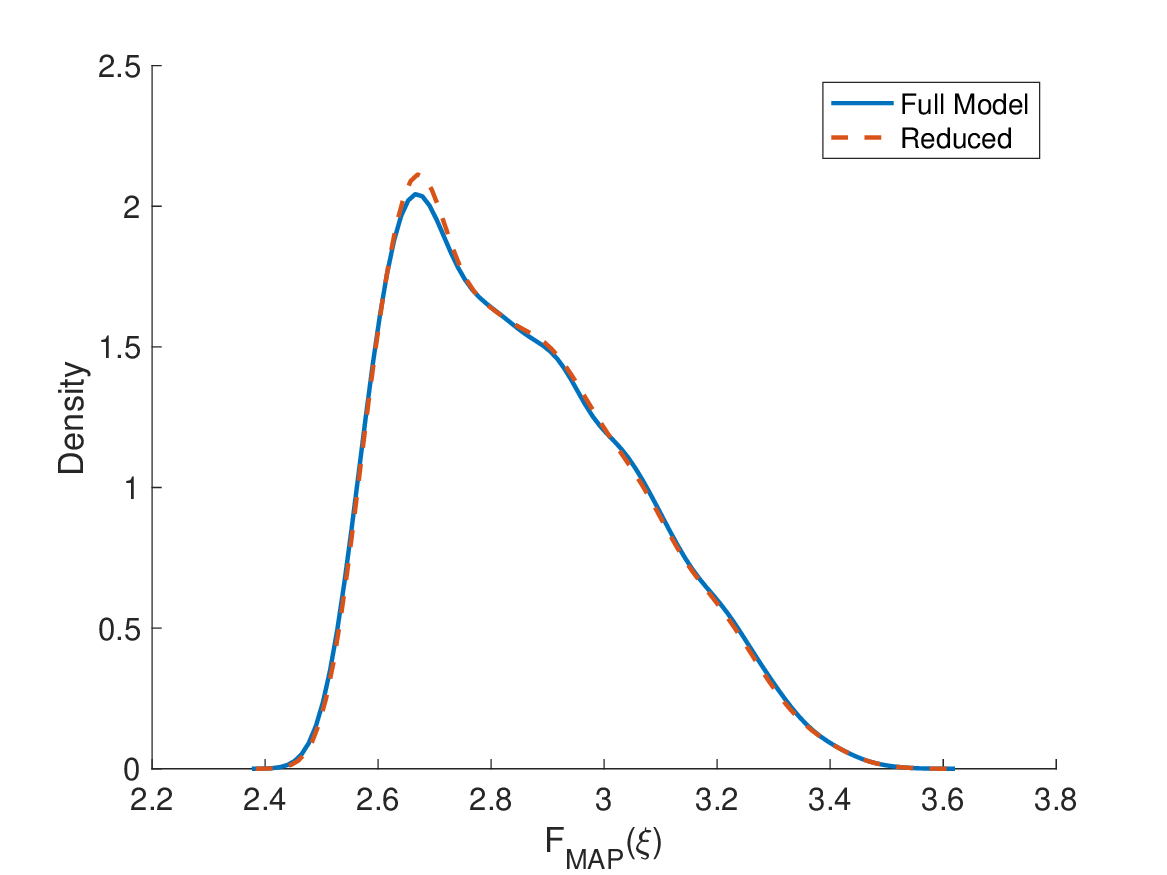}} 
    \caption{Kernel density estimates of $F(\boldsymbol\xi)$ when sampled over
    all prior hyperparameters compared to when the prior mean of $\log\gamma$
    and prior variances of $\log\mu,\log\gamma$ are fixed at the nominal values 
    of -1.5, 1, and 1, respectively.
    These estimates on superimposed on one another for
    $F_\mathrm{mean}(\boldsymbol\xi)$ (left), $F_\mathrm{var}(\boldsymbol\xi)$
    (right), and $F_\mathrm{MAP}(\boldsymbol\xi)$ (bottom).}
    \label{fig:seir_red}
\end{figure}

Overall, we note that the results from the SW-ELM and sparse regression PCE
results agree.  The indicates that the present computations are stable with
respect to the choice of the surrogate model. 
The global sensitivity
analysis of the posterior mean, variance, and MAP point in~\cref{fig:seir_bar}
allow us to infer much information about which hyperparameters in the prior
matter and which do not.  The Sobol' indices suggest that the uncertainty in 
the prior mean of
$\log\gamma$ and prior variances of $\log\mu,\log\gamma$ can be ignored.
To illustrate this, 
we compare the distributions of $F_\mathrm{mean}, F_\mathrm{var}$, and
$F_\mathrm{MAP}$ before and after these prior hyperparameters are fixed
at their nominal values
in~\cref{fig:seir_red}.
The density estimates in~\cref{fig:seir_red} confirm that those three prior
hyperparameters have little influence over the posterior mean, variance, and MAP
point. Thus, the experimental resources should be put towards finding more
knowledge about the other hyperparameters. 


\section{Conclusion}
We have developed a computational approach for global sensitivity analysis of
Bayesian inverse problems with respect to hyperparameters defining the prior.
Our results indicate that the posterior distribution can exhibit complex
dependence on such hyperparameters.  
Consequently, the uncertainty in the prior hyperparameters lead to 
uncertainty in posterior statistics of the prediction/goal quantities of
interest which needs to be accounted for. 
The results of GSA provide valuable
insight this context.  Such an analysis reveals the prior
hyperparameters that are most influential to the posterior statistics of
prediction quantities of interest and whose specification requires care. 
Our computational studies provide a
proof-of-concept of the proposed approach and indicate its viability.  In
particular, at the cost of one MCMC run, we can obtain reliable estimates of the
sensitivity of moment-based hyperparameter-to-statistic mappings with respect to
prior hyperparameters.

An important aspect of our approach is the proposed importance sampling
procedure.  A limitation of the present study is that the importance sampling
prior in~\eqref{equ:pipostIS} was chosen in an empirical manner. While this can
be practical in many cases, developing a systematic approach for picking this
distribution is an interesting and important avenue of future investigations.
This can be facilitated, e.g., by considering an appropriate optimization
problem for finding $\pipris$.  This requires definition of suitable performance
objectives for $\pipris$ that are tractable to optimize.  


A related line of inquiry  is exploration of techniques such as variational
inference~\cite{VarInf17} or the Laplace approximation~\cite{BDA1} to the
posterior for obtaining an importance sampling posterior $\pipostis$.  This is
necessary for computationally intensive inverse problems where even one MCMC run
might be prohibitive.  Yet another direction for future work is the development of
hyperparameter screening steps. A tried-and-true approach is to screen via derivative-based global sensitivity
measures~\cite{KucherenkoSobol09,KucherenkoIooss17}, after which a variance-based analysis may be conducted.  This would be
important for inverse problems with a large number of prior hyperparameters.


\bibliographystyle{elsarticle-num}
\bibliography{refs}


\end{document}